\documentclass[fleqn,usenatbib]{mnras}

\usepackage{newtxtext,newtxmath}

\usepackage[T1]{fontenc}

\DeclareRobustCommand{\VAN}[3]{#2}
\let\VANthebibliography\thebibliography
\def\thebibliography{\DeclareRobustCommand{\VAN}[3]{##3}\VANthebibliography}

\usepackage{amsmath,amstext} 
\usepackage[pdftex]{graphicx} 
\usepackage{caption}
\usepackage{subcaption}

\graphicspath{ {./images/} }

\newcommand{\sbr}[1]{_{\mathrm{#1}}}
\newcommand{\spr}[1]{^{\mathrm{#1}}}
\newcommand{\ecrit}{\Sigma\sbr{crit}}

\newcommand{\Msun}{\mbox{M}\ensuremath{_{\odot}}}
\newcommand{\invEcrit}{\langle\Sigma\sbr{crit}^{-1}\rangle^{-1}}

\newcommand{\wpm}{w\sbr{p}\spr{m}}

\title[Shape of Dark Matter Haloes]{The shape of dark matter haloes: results from weak lensing in the Ultraviolet Near-Infrared Optical Northern Survey (UNIONS)}

\author[B. Robison et al.]{%
Bailey Robison$^{1,2}$\thanks{E-mail: baileyarobison@gmail.com},
Michael J. Hudson$^{1, 2, 3}$, Jean-Charles Cuillandre$^{4}$, Thomas Erben$^{5}$, \newauthor S\'ebastien Fabbro$^{6}$, Rapha\"el Gavazzi$^{7, 8}$, Axel Guinot$^{9, 10}$, Stephen Gwyn$^{11}$, \newauthor Hendrik Hildebrandt$^{12}$, Martin Kilbinger$^{9, 13}$, Alan McConnachie$^{6}$, Lance Miller$^{14}$, \newauthor Isaac Spitzer$^{1, 2}$, Ludovic van Waerbeke$^{15}$
\\
$^{1}$Department of Physics and Astronomy, University of Waterloo, Waterloo, ON, N2L 3G1, Canada\\
$^{2}$Waterloo Centre for Astrophysics, Waterloo, ON, N2L 3G1, Canada\\
$^{3}$Perimeter Institute for Theoretical Physics, 31 Caroline St. N., Waterloo, ON, N2L 2Y5, Canada\\
$^{4}$ AIM, CEA, CNRS, Université Paris-Saclay, Université de Paris, F-91191, Gif-sur-Yvette, France\\
$^{5}$Argelander-Institut f\"ur Astronomie, Auf dem H\"ugel 71, 53121 Bonn / Germany\\
$^{6}$NRC Herzberg Astronomy \& Astrophysics, 5071 West Saanich Road, British Columbia, Canada V9E2E7\\
$^{7}$ Sorbonne Université, CNRS, UMR7095, Institut d’Astrophysique de Paris, F-75014, Paris, France\\
$^{8}$ Institute of Astronomy, University of Cambridge, Madingley Road, Cambridge CB30HA, UK\label{ioa}\\
$^{9}$AIM, CEA, CNRS, Université Paris-Saclay, Université de Paris, 91191 Gif-sur-Yvette, France\\
$^{10}$Université de Paris, CNRS, Astroparticule et Cosmologie, F-75013 Paris, France\\
$^{11}$Canadian Astronomy Data Centre, Herzberg Astronomy and Astrophysics, National Research Council, 5071 West Saanich Rd, Victoria BC, V9E 2E7\\
$^{12}$ Ruhr-Universität Bochum, Astronomisches Institut, German Centre for Cosmological Lensing (GCCL), Universitätsstr. 150, 44801, Bochum, Germany\\
$^{13}$Institut d’Astrophysique de Paris, UMR7095 CNRS, Université Pierre \& Marie Curie, 98 bis boulevard Arago, 75014 Paris, France\\
$^{14}$Department of Physics, University of Oxford, Denys Wilkinson Building, Keble Road, Oxford OX1 3RH, UK\\
$^{15}$Department of Physics and Astronomy, University of British Columbia, 6224 Agricultural road, Vancouver, BC V6T 1Z1, Canada\\
}

\date{Accepted XXX. Received YYY; in original form ZZZ}

\pubyear{2022}

\begin{document}
\label{firstpage}
\pagerange{\pageref{firstpage}--\pageref{lastpage}}
\maketitle

\begin{abstract}
Cold dark matter haloes are expected to be triaxial, and so appear elliptical in projection. We use weak gravitational lensing from the Canada-France Imaging Survey (CFIS) component of the Ultraviolet-Near Infrared Optical Northern Survey (UNIONS) to measure the ellipticity of the dark matter haloes around Luminous Red Galaxies (LRGs) from the Sloan Digital Sky Survey Data Release 7 (DR7) and from the CMASS and LOWZ samples of the Baryon Oscillation Spectroscopic Survey (BOSS), assuming their major axes are aligned with the stellar light. We find that DR7 LRGs with masses $M \sim 2.7\times10^{13} \Msun/h$ have halo ellipticities $e=0.46\pm0.10$. Expressed as a fraction of the galaxy ellipticity, we find $f_h = 2.2\pm0.6$. For BOSS LRGs, the detection is of marginal significance: $e = 0.20\pm0.10$ and $f_h=0.7\pm0.7$. These results are in agreement with other measurements of halo ellipticity from weak lensing and, taken together with previous results, suggest an increase of halo ellipticity of $0.10\pm0.06$ per decade in halo mass. This trend agrees with the predictions from hydrodynamical simulations, which find that at higher halo masses, not only do dark matter haloes become more elliptical, but that the misalignment between major axis of the stellar light in the central galaxy and that of the dark matter decreases.
\end{abstract}

    


\section{Introduction}

In the standard $\Lambda$CDM cosmological model of our universe, dark matter accounts for over 80\% of the matter content and plays a dominant role in the formation and evolution of large scale structure. Dark matter haloes, in which galaxies reside, assemble in a hierarchical manner, with less massive haloes accreting onto more massive ones. Simulations have revealed that filaments and other forms of large scale structure will have an effect on the rate and direction of the accretion of smaller haloes \citep{VanVan93}, and that this leads to triaxial dark matter haloes that appear elliptical in projection \citep{DubCar91,JingSuto02, BaiSte05}, with more massive haloes tending to be more elliptical \citep{AllPriKra06}.

This halo anisotropy has implications for cosmological studies with weak lensing. Weak lensing operates under the assumption that galaxies are randomly oriented, while in reality they will have alignments due to the gravity of surrounding structure. Therefore, this intrinsic alignment is a significant source of contamination. Improving our model of halo anisotropy will allow us to develop a better understanding of these intrinsic alignments, which will improve the results of future weak lensing studies. Dark matter halo anisotropy also provides a test to rule out theories of modified gravity \citep{Milgrom13,Khoury15}, and provides a method to constrain the cross section of self-interacting dark matter \citep{DavSpeSte01,PetRocBul13}.

The distribution of satellite galaxies has been used to infer the shape of dark matter haloes. Studies have focused on the distribution of satellite galaxies with respect to the shape of the central galaxy \citep{Brainerd05,AzzPatPra07}. A preferential alignment of the satellite distribution with the major axis of the light of the central galaxy is evidence of a non-spherical dark matter distribution. This preferential alignment has been confirmed by observations \citep{YangVanMo06} and in simulations \citep{ZenKraGne05,LibColFre07}.

Weak lensing is another method of detecting dark matter halo anisotropy \citep{SchneiderBartelmann1997,  NatRef00, BrainerdWright2000}. This involves  measuring the azimuthal dependence of the shear. \cite{NatRef00} proposed splitting the weak lensing shear into monopole and quadrupole terms, where the quadrupole is aligned with the major axis of the galaxy's light. \cite{HoeYeeGla04} introduced the measurement of  $f_h$, the ratio of the aligned ellipticities of haloes and galaxy light, and measurements of $f_h$ have been made by \cite{Mandelbaum06} and \cite{Schrabback15}. Recently, this method has been used by \cite{Schrabback21} to obtain a 3.8$\sigma$ detection of halo ellipicity, one of the most significant detections of anisotropy of galaxy-scale haloes. \cite{BrainerdWright2000} proposed comparing the weak lensing signal within $45\deg$ of the major and minor axes, a method implemented by \cite{ParHoeHud07}. This method was also utilised by \cite{VanHoeJoa17} to obtain a significant detection of halo ellipticity in group-scale haloes. The method was extended and improved by \cite{ClaJai16} who found a $4\sigma$ detection of the halo ellipticity of luminous red galaxies (LRGs) from the Sloan Digital Sky Survey (SDSS). This method has also been used to measure the halo ellipticity of massive galaxy cluster haloes \citep{EvaBri09, OguTakOka10}, where the lensing is much stronger.

In this paper, we present a new measurement of halo ellipticity from weak lensing.
In Section~\ref{sec:data}, we present the source and lens data used in this paper. 
In Section~\ref{sec:analysis} we explain our methods of analysis, including the anisotropic halo model and the various estimators used to measure quadrupole shear.
We present our results in Section~\ref{sec:results}. These include results from the monopole shear and the average halo ellipticity from the quadrupole shear. 
A discussion of systematic effects is given in Section~\ref{sec:systematics}. 
Section~\ref{sec:discussion} includes a comparison with previous results. Finally, in Section~\ref{sec:conclusion} we present our conclusions and prospects for future results.

The adopted cosmology is a flat Universe with $\Omega\sbr{0} = 0.3$ and we quote all factors that depend on $H_0$ using $h \equiv H_0/(100$ km/s/Mpc). There are two conventions for galaxy (or halo) ``ellipticity'', and in this paper, for comparison with the literature, we use both. We adopt the following notation: $e \equiv \frac{a^2-b^2}{a^2+b^2}$ and $\epsilon \equiv \frac{a-b}{a+b}$, where $a$ and $b$ are the major and minor axes, respectively.

\section{Data}
\label{sec:data}

\subsection{Source galaxies from UNIONS}

The sources used in our weak lensing analysis were derived from the $r$-band component of the  Ultraviolet Near-Infrared Optical Northern Survey (UNIONS)\footnote{https://www.skysurvey.cc/}. UNIONS is a deep wide-field multi-band ($ugriz$) imaging survey covering the high Galactic latitude sky north of 30 degrees declination (approximately 4,800 square degrees). The $r$-band component of UNIONS was obtained at the Canada-France-Hawai`i Telescope as part of the Canada-France Imaging Survey \citep[CFIS,][]{IbataMcConnachieCuillandre2017} with a median seeing of 0.68 arcseconds and a 10-$\sigma$ limiting magnitude of 24.1 for extended sources in the $r$-band.

The UNIONS $r$-band data used in this paper cover 1,565 square degrees of the northern hemisphere, containing roughly 46 million source galaxies. This coverage can be divided into 4 contiguous regions, referred to as ``patches''. To save computer time and memory, the weak lensing analysis is performed on each of these individual patches rather than on the entire source catalogue simultaneously. The patches are large enough that any effects from being near the edge of the source catalogue are minimal. Patch 1 is the largest, containing nearly half of the sources. Patch 4 contains roughly a quarter of the sources, while Patches 2 and 3 contain roughly an eighth.

Galaxy shape measurements, necessary for weak lensing, are generated using an early version of \textsc{ShapePipe}, a new shape measurement pipeline \citep{Guinot22}. The pipeline uses the \texttt{ngmix} package \citep{Sheldon15} to perform {\sc MetaCalibration} \citep{HufMan17} which yields the ellipticities $\epsilon_1$ and $\epsilon_2$ for each source. Each source is assigned a statistical weight that quantifies how well the image is fit by the resulting shape. The weight is given by
\begin{equation}
    w = \frac{1}{2\sigma\sbr{int}^2 + \sigma_{\epsilon_1}^2 + \sigma_{\epsilon_2}^2} \,
\end{equation}
The intrinsic shape noise is $\sigma\sbr{int} = 0.34$ for both components  \citep{Guinot22}. The two parameters $\sigma_{\epsilon_i}^2$ are the variances of measurement errors on the ellipticities.

To convert the shear into a mass distribution, we need the critical surface mass density \begin{equation}
    \ecrit = \frac{c^2}{4\pi G} \frac{D\sbr{s}(z\sbr{s})}{D\sbr{l}(z\sbr{l}) D\sbr{ls}(z\sbr{l}, z\sbr{s}) } \,,
\label{eq:sigma_crit}
\end{equation}
where the distances are angular diameter distances that depend on the lens or source redshifts, or both. At present, UNIONS data do not have complete deep \emph{ugriz} photometry, and photometric redshifts of the source galaxies are not yet available. Nevertheless we can make a statistical determination of the critical density if the source redshift distribution is known. We have measured $p(z\sbr{s})$ using the method described in \cite{LimCunOya08}, with the implementation of \cite{HilVioHey17, HilKohVan20}. Full details of the application of this method to UNIONS are given in Spitzer et al. (submitted); here we give a brief summary. We match the UNIONS catalogue with the W3 patch of the deeper Canada-France-Hawaii Lensing Survey (CFHTLenS), which overlaps with UNIONS to obtain \textit{ugriz} photometry for CFIS sources. Then, a spectroscopic sample, with \textit{ugriz} photometry from CFHTLenS, was used \citep{HilErbKui12, ErbHilMil13}, and galaxies in this sample were reweighted until their distribution in 5-dimensional colour space matched that of the \textit{ugriz} UNIONS catalogue. The resulting reweighted $p(z)$ from the sample was then adopted as the $p(z)$ for the UNIONS catalogue. The resulting $p(z)$ is fit by the profile described in equation~(\ref{eq:source_redshift}). We also create versions of the catalogue that are magnitude limited and fit with the same profile to account for varying depth across the field. The distribution of parameters as a function of weighted median \textit{r}-band magnitude is fit by equations~(\ref{eq:am}) and (\ref{eq:zm}). 
We fit the data with a source redshift function of
\begin{equation}
    p(z) = \left( A(m)\cdot \frac{z^{\alpha}\cdot e^{-\left( \frac{z}{z_0(m)} \right)^{\alpha}}}{\frac{z_0^{\alpha+1}}{\alpha}\cdot \Gamma\left( \frac{\alpha+1}{\alpha} \right)} \right)
    +
    \left( (1-A(m))\cdot \frac{e^{-\frac{(z-\mu)^2}{2\sigma^2}}}{2.5046} \right) \,,
\label{eq:source_redshift}
\end{equation}
\begin{equation}
    A(m) = -0.4154 m^2 + 19.1734 m - 220.261 \,,
\label{eq:am}
\end{equation}
\begin{equation}
    z_0(m) = 0.1081 m - 1.9417 \,,
\label{eq:zm}
\end{equation}
where $\alpha = 1.79$, $\sigma = 1.3$, $\mu = 1$ and where $m$ is the median weighted $r$-band magnitude. This median $r$-band magnitude is different for each of the 4 patches, so each patch will use a slightly different $p(z)$ and $\ecrit^{-1}(z\sbr{l})$. The median magnitude for Patch 1 is $m = 22.95$, for Patch 2 it is $m = 22.944$, for Patch 3 it is $m = 22.923$, and for Patch 4 it is $m = 22.952$. The process of fitting $p(z)$ was undertaken by Spitzer et al. (submitted).

We then calculate the average \emph{inverse} critical density, following \cite{VioCacBro15}, by integrating over the probability density function of the source galaxy redshifts as follows
\begin{equation}
    \langle\ecrit^{-1}(z\sbr{l})\rangle = \frac{4\pi G}{c^2} \int_{z\sbr{l}}^{\infty} \frac{D\sbr{l}(z\sbr{l})D\sbr{ls}(z\sbr{l},z\sbr{s})}{D\sbr{s}(z\sbr{s})} p(z\sbr{s})dz\sbr{s} \,.
\label{eq:sigma_crit_func}
\end{equation}
This can be evaluated for each lens and incorporated into the weighted average to obtain the excess surface mass density (ESD), $\Delta\Sigma$.

We note, however, that the halo ellipticity is robust to the source redshift distribution because it depends on a ratio of the quadrupole moment of the ESD to the monopole of the ESD. Therefore, any systematic error in the ESD, due to, for example, a systematic error in the source redshift distribution, will appear in both numerator and denominator, and hence will cancel. 

\subsection{Lens Galaxies}
\label{sec:lens_data}

Luminous Red Galaxies (LRGs) are used as lenses because they reside in massive dark matter haloes, with a typical halo mass on the order of $10^{13}-10^{14}h^{-1}$M$_{\odot}$ \citep{ZehZehEis09}. A more massive halo has a stronger weak lensing signal, which makes it easier to detect the quadrupole component of the shear. Also, simulations suggest that more massive haloes tend to be more elliptical \citep{AllPriKra06}. Finally, LRGs provide a reliable method of aligning our lensing measurements. The distribution of satellite galaxies, which may trace the dark matter halo, are more aligned with the galaxy light for red central galaxies \citep{YangVanMo06}.

We consider two LRG samples in this paper. One lens sample consists of LRGs from the SDSS DR7 catalogue of \cite{KazinBlantonScoccimarro2010}. These LRGs span a redshift range of $0.15 < z < 0.5$ with a median redshift $z=0.34$ and a median lens galaxy ellipticity of $e=0.22$. This lens sample has $\langle\Sigma\sbr{crit}^{-1}\rangle^{-1}=6120$ M$_{\odot}$h/pc$^2$. Only LRGs that overlap with the current UNIONS weak lensing footprint were used, resulting in a lens sample of approximately 18,000 LRGs. 

The second lens sample consists of the LRGs from the CMASS and LOWZ samples of the BOSS component of SDSS-III \citep{DawsonSchlegelAhn2013}. These LRGs are selected within several magnitude and colour criteria, and span a redshift range of $0.15 < z < 0.7$. Only LRGs that overlap with the current UNIONS footprint were used. This resulted in a lens sample of approximately 144,000 lenses with a median redshift of $z = 0.51$ and a median lens galaxy ellipticity of $e=0.22$. This lens sample has $\langle\Sigma\sbr{crit}^{-1}\rangle^{-1}=8313$ M$_{\odot}$h/pc$^2$.

In order to measure the anisotropy of the shear signal, the lenses need to be aligned before stacking.  We first matched the SDSS LRG catalogue with UNIONS photometric catalogues on position. Many LRGs are too large or bright for our standard shape measurement pipelines, which are designed for small faint galaxies near the survey limit. Instead, the shape and orientation of LRG in UNIONS was obtained from the position angles and axis ratios in the UNIONS $r$-band SExtractor \citep{BerArn96} catalogue. The SExtractor position angles are not corrected for PSF anisotropy, but the source galaxy shapes are. So in principle, there should be no correlation between shapes of lenses and sources. Nevertheless, as a test, we also perform the analysis with LRG position angles derived from independent SDSS photometry in Section \ref{sec:systematics}.

\subsection{The effect of satellites of lens galaxies}
\label{sec:satellites}
There is no photometric redshift information for UNIONS source galaxies yet. As a result, we have no way of knowing which sources are actually behind the lens and therefore which are affected by weak lensing. Some portion of the sources surrounding the lenses are actually satellite galaxies. These satellite galaxies are not affected by lensing, and may introduce bias in the form of coherent intrinsic alignment. \cite{SchBri10} propose a model for intrinsic alignments of galaxies. Their model is based on the linear alignment model, which assumes the alignment of galaxies is linearly proportional to the tidal field. This model predicts preferential radial alignment of satellites with the central galaxy. However, observational studies have found mixed results. Several studies of satellite alignment yield results consistent with random alignment \citep{SifHoeCac15, SchColFre13}, although preferential radial alignment has been observed in others \citep{SinManMor15,Georgiou19}.

We also expect there to be an excess of satellites along the major axis of the lens light and a deficit along the minor axis, which will influence our measurement of the elliptical shear signal. This alignment between the satellite distribution and the alignment of the central galaxy has been observed by \cite{YangVanMo06}, who found a stronger alignment for red central galaxies. Our lenses, being LRGs, are therefore expected to be significantly aligned with their satellite populations. 

Light from the lens will contaminate the shape measurements of the plentiful sources along the lens major axis. This can lead to the shape measurement being biased in the radial direction \citep{SifHerHoe18}. If there is a significant preferential radial alignment, an excess of satellites along the major axis will lead to a larger negative contribution near the major axis. This will have the effect of adding a negative quadrupole term, or a negative halo ellipticity, to our results.

\begin{figure}
\includegraphics[width=\linewidth]{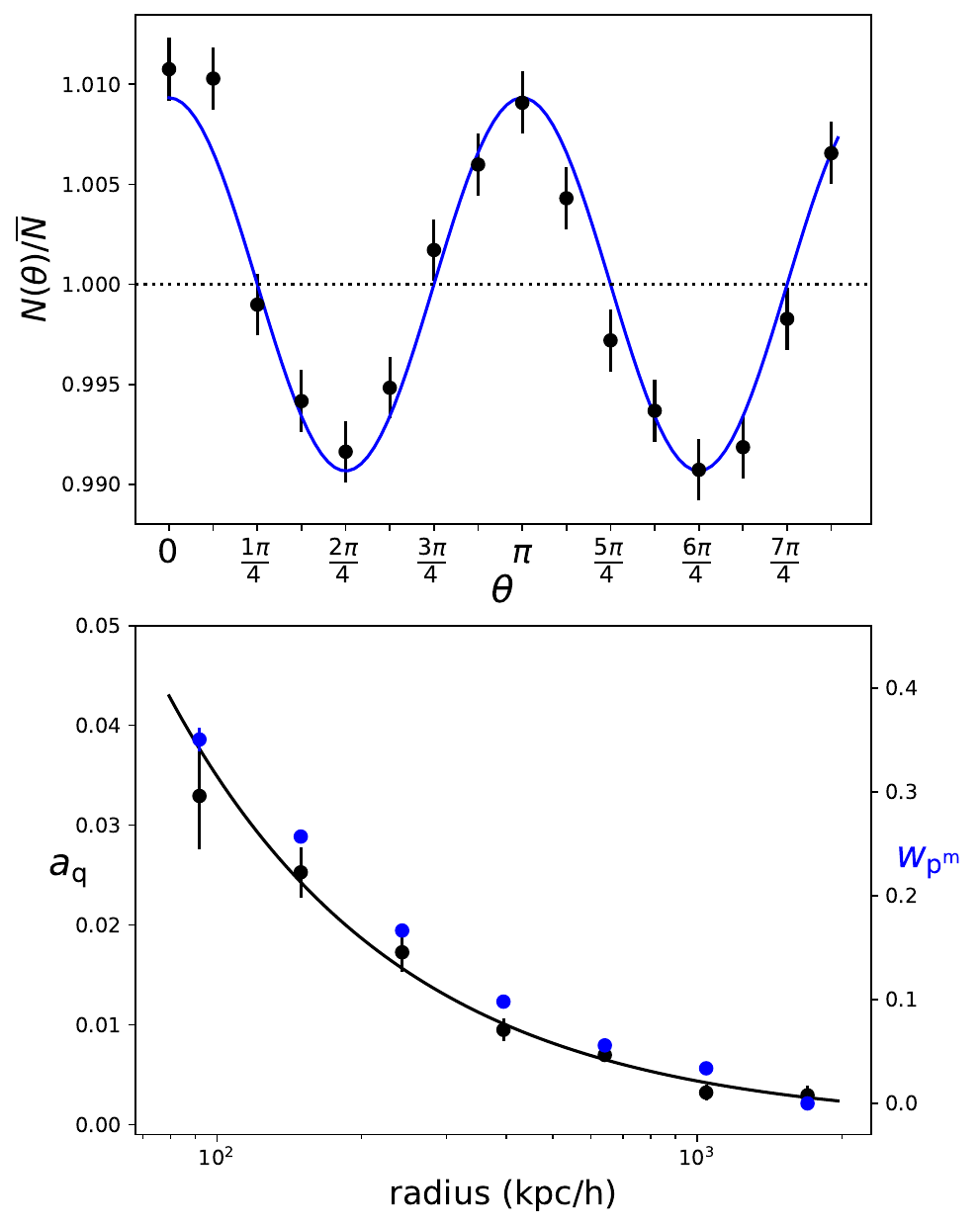}
\caption{Top: Weighted source galaxy counts in 16 azimuthal bins. The horizontal axis the angle from the major axis of the DR7 LRG's stellar light. Only source galaxies with a projected separation of 110-500 kpc/h are included. A function of the form $\cos(2\theta)$ was fit to determine the amplitude. The vertical axis is scaled to show deviation from the mean. Bottom: Amplitude fit to azimuthal source distribution as a function of radius (black circles). The right-hand scale and the blue squares show the excess counts above background.}
\label{fig:angular_bins}
\end{figure}

In order to account for the effect of satellite galaxies on our results (which will be described in more detail in Section \ref{sec:model_satellites}),  we need to model the anisotropic distribution of satellites. We write the observed surface number density of source galaxies around a lens galaxy as  
\begin{equation}
    \frac{n(R,\theta)}{n\sbr{b}} =  1  + w\sbr{p}(R, \theta) = [1  + w\sbr{p}\spr{m}(R)][ 1 + a\sbr{q} (R) \cos(2\theta) ]\,,
    \label{eq:wmp_and_aq}
\end{equation}
where $n\sbr{b}$ is the surface number density of background source galaxies (assumed to be uniform), and $w\sbr{p}(R, \theta)$ is the anisotropic projected cross-correlation of sources with a lens galaxy. In the second equality, we separate this into a monopole excess $\wpm$ and a quadrupolar angular dependence,  where $\theta = 0 $ is aligned with the major axis of the light, and the amplitude of the quadrupole is denoted $a\sbr{q}$. 

To measure these quantities, the lenses were rotated to a coordinate system where the major axis of the lens galaxy light (as measured in the UNIONS catalogue) is aligned with the $x$-axis. The region within 140-4200 kpc/$h$ from each LRG was divided into concentric annular bins. In each of those radial bins weighted source galaxy counts were binned by azimuthal angle from the major axis. For each angular bin, the uncertainty is the square root of the number of sources in the bin.

For each radial bin we calculate $\wpm$, which is defined as the ratio of the number density of sources within each bin to the number density of background sources. We also calculate the amplitude of quadrupole, $a\sbr{q}$. To do so, in each radial bin, we fit a function of the form $\cos(2\theta)$ to the angular source distribution. An example of this fit for DR7 LRGs is shown in the upper panel of Fig.~\ref{fig:angular_bins}. Sources within 110 -- 500 kpc/h were divided into 16 angular bins, and a fit was performed to determine the amplitude $a\sbr{q}$. This was repeated for each radial bin to obtain $a\sbr{q} (R)$.

The bottom panel displays $a\sbr{q} (R)$ in black and $\wpm (R)$ in blue for several radial bins. A power law was fit to each. We will use these power law fits to interpolate $\wpm$ and $a\sbr{q}$ at any distance from the center of the lens. 
This process was repeated for the BOSS lenses, and for the DR7 lenses but taking the lens major axes angles from SDSS (see Section \ref{sec:systematics}).

This radial dependence in the alignment of the satellite distribution is in agreement with other works \citep{YangVanMo06}. If satellites trace the halo, this suggests that the dark matter halo is well aligned with the lens light. Therefore, we use the lens light as a proxy for the major axis of the halo when stacking to measure the quadrupole shear. In addition to their usefulness as a possible proxy for shape of the dark matter haloes, satellite galaxies are also a potential source of contamination for our shear measurements. Satellite galaxy orientations are expected to be preferentially aligned with the central galaxy \citep{SchBri10, Georgiou19}, which could lead to a radial bias in the shear.

\cite{Georgiou19} observe a radial dependence on satellite alignment with respect to their group's BCG. Satellites close to the BCG experience a radial alignment, which affects the tangential component of their shape. They observe no effect on the cross component. In order to measure this radial contamination, we fit a power law to Figure 3 in \cite{Georgiou19}. We omit the closest radial bin from our fit due to potential contamination from the BCG light. This radial contamination uses the ellipticity notation $\epsilon$, which is the same notation used in our source shape measurements. This contamination, $\epsilon\spr{int}$, can be combined with $\invEcrit$ to obtain the contamination in the mass, $\Delta\Sigma\sbr{int}$. We will discuss the impact of this on our measurements in Sections \ref{sec:model_satellites} and \ref{sec:quad_estimators}.

\section{Analysis}
\label{sec:analysis}
Before measuring the halo ellipticity, we first determine the mass and concentration of the LRG halo by measuring the monopole component of the tangential shear. This is commonly calculated as a weighted average of the source galaxy ellipticities. All sources have a weight, $w\sbr{s}$, that describes the quality of the source's shape measurement. We also weight each lens-source pair by $W\sbr{l}=\langle\ecrit^{-1}(z\sbr{l})\rangle^2$ following \cite{SheJohFri04}, with $\langle\ecrit^{-1}(z\sbr{l})\rangle$ from equation~(\ref{eq:sigma_crit_func}). The excess mass density is given by
\begin{equation}
\langle\Delta\Sigma(R)\rangle = \frac{\sum \epsilon\sbr{+,s}\,\langle\ecrit^{-1}(z\sbr{l})\rangle^{-1} \, w\sbr{s} W\sbr{l}}{\sum w\sbr{s} W\sbr{l}} \,,
\label{eq:sigmastack}
\end{equation}
summing over all sources, s, and all lenses, l, in a given radial separation. 

After measuring the monopole shear, we measure the quadrupole shear. This process is different, as we need to take the orientation of the lenses into account. First, the positions of the lens and the sources are converted from equatorial coordinates into a local 2D Cartesian coordinate system centred on the lens. All data, including the positions and shapes of the sources, are rotated so that the major axis of the light in the lens is aligned with the $x$-axis of the coordinate system. This process is repeated for each lens to measure the azimuthal variation in the shear and the results are stacked.

\subsection{Model}
Our model, which relates the measured shear to the mass and ellipticity of the dark matter halo, consists only of the so-called ``1-halo'' term, which describes the matter directly attached to the lens galaxy. One could also include an ``offset group'' term to account for lenses that reside within subhaloes inside a larger host halo. However, few of the LRGs are expected to be satellite galaxies: for example, from halo occupation modelling, \cite{ZhengZehaviEisenstein2009} predict satellite fractions of 2-5\% (depending on luminosity) for DR7 LRGs. The satellite fraction for the BOSS LRGs, which are less massive than the DR7 LRGs, is somewhat higher but still low: about 10\% \citep{WhiteBlantonBolton2011, ParejkoSunayamaPadmanabhan2013}. Consequently we neglect this term in the modelling. A 2-halo term is often included to account for the lensing signal from neighbouring haloes. This term is important at large radii, but not within the radial range we are concerned with in this work.

The 1-halo term is comprised of the contribution from the stellar mass of the galaxy and the mass of the galaxy's dark matter halo. The stellar mass is treated as a point mass, 
\begin{equation}
\Delta\Sigma_*(R) = M_*/\pi R^2 \,.
\label{eq:stellar_mass}
\end{equation}
We also require a term to describe the dark matter halo in the 1-halo term. Normally, an NFW profile is used to describe the density profile of a dark matter halo. However, the mass of the NFW profile does not converge when integrated to an infinite radius. Instead, we use a truncated NFW profile from \cite{BalMarOgu09} to describe the dark matter halo, which has a well defined total mass. The truncated NFW profile is
\begin{equation}
\rho(x)=\frac{M_0}{4\pi r_s^3}\frac{1}{x(1+x)^2}\frac{\tau^2}{\tau^2+x^2} \,,
\end{equation}
where $x=r/r_s$ and $M_0=4\pi \rho_s r_s^3$. The truncation factor, $\tau=r_t/r_s$, describes the radius where the truncation term begins to dominate. For the remainder of this paper, a truncation factor of $\tau=10$ is used. The scale radius of the halo, $r_s$, is related to the virial radius ($R_{200c}$) through the concentration ($c_{200c}$). The scale density, $\rho_s$, is related to the critical density and also depends on the concentration.
\begin{equation}
\rho_s = \frac{200}{3}\frac{c_{200c}^3}{\ln(1+c_{200c}) - \frac{c_{200c}}{1+c_{200c}}}\rho_c
\label{eq:scale_density}
\end{equation}
We parametrise the NFW profile using only $M_{200c}$ and $c_{200c}$. The total circularly symmetric model  comprised of this truncated NFW profile and the stellar mass term from equation~(\ref{eq:stellar_mass}):
\begin{equation}
\Delta\Sigma\sbr{m} = \Delta\Sigma_* + \Delta\Sigma\sbr{NFW}
\end{equation}

Now we can extend this model to account for the halo ellipticity, following \cite{Adhikari2015} and \cite{ClaJai16}. The surface mass density of the anisotropic halo can be split into a monopole, described by a projection of the truncated NFW profile, and a quadrupole term:
\begin{equation}
\Sigma(R,\theta) = \Sigma\sbr{m}(R)\left[ 1 - \frac{e}{2}\eta(R)\cos(2\theta) \right] \,,
\label{eq:aniso_model}
\end{equation}
where $\theta$ is defined as the angle measured counter-clockwise from the major axis of the mass distribution. The ellipticity of the halo is represented by $e$. The function $\eta(R)$ describes how the quadrupole term is related to the monopole term at different radii. \cite{ClaJai16} use
\begin{equation}
\eta(R) = \frac{\text{d}\log\Sigma\sbr{m}(R)}{\text{d}\log R} = \frac{R}{\Sigma\sbr{m}(R)} \frac{\text{d}\Sigma\sbr{m}(R)}{\text{d}R} \,,
\label{eq:eta_0}
\end{equation}

It is also useful to separate the tangential shear into a monopole term and a quadrupole term:
\begin{equation}
 \gamma\sbr{+}(R, \theta) = 
 \gamma\sbr{+}\spr{m}(R) + \gamma\spr{q}\sbr{+}(R) \cos(2\theta) \,.
\end{equation}
The cross-shear has no monopole, and its quadrupole is 
\begin{equation}
 \gamma\sbr{\times}(R, \theta) =  \gamma\spr{q}\sbr{\times}(R)
 \sin(2\theta) \,.
\end{equation}
\cite{Adhikari2015} found that the tangential and cross-components of shear from the quadrupole are given by
\begin{equation}
\begin{aligned}
& \Sigma_{\text {crit }} \gamma\sbr{+}\spr{q}=(e / 2)\left[\Sigma\sbr{m}(R) \eta(R)-I_1(R)-I_2(R)\right] \\
& \Sigma_{\text {crit }} \gamma\sbr{\times}\spr{q}=(e / 2)\left[-I_1(R)+I_2(R)\right]
\end{aligned}
\label{eq:adhikari_model}
\end{equation}
where
\begin{equation}
\begin{aligned}
I_1(R) & \equiv \frac{3}{R^4} \int_0^R R^{\prime 3} \Sigma\sbr{m}\left(R^{\prime}\right) \eta\left(R^{\prime}\right) \mathrm{d} R^{\prime} \\
I_2(R) & \equiv \int_R^{\infty} \frac{\Sigma\sbr{m}\left(R^{\prime}\right) \eta\left(R^{\prime}\right)}{R^{\prime}} \mathrm{d} R^{\prime}
\end{aligned}
\end{equation}

\subsection{Allowing for contamination by satellite galaxies}
\label{sec:model_satellites}
As discussed in Section~\ref{sec:lens_data}, we need to account for the presence of satellite galaxies. These satellites have two effects. First, they dilute the background source galaxies. Second, due to tidal fields, they are radially aligned with the host halo (a form of ``intrinsic alignments'')  leading to an underestimation in the weak lensing signal. We need to adjust our model to account for these effects.

Now suppose that the satellites associated with the lens are intrinsically aligned with the lens with some tangental ``shear'' (more accurately, ellipticity) $\epsilon\sbr{+}\spr{int}$. As discussed, recent results \citep{Georgiou19} have shown that these intrinsic alignments are radial. Since we adopt the convention that a tangential shear is positive, $\epsilon\sbr{+}\spr{int}$ is negative.

The average tangential shear that one would predict for a sample with satellites is a weighted average of  lensing tangential shear of background sources, contaminated and diluted by satellites in the lens galaxy's anisotropic halo:
\begin{equation}
\overline{\gamma}(R,\theta) = \frac{\gamma(R) + w\sbr{p}(R,\theta) \epsilon\sbr{+}\spr{int}(R)}{1 + w\sbr{p}(R, \theta)} \,.
\end{equation}
To simplify the notation let us drop the $R$ and write the $\theta$ dependence explicitly as a quadrupole, separating these into $\cos(2\theta)$ and $\sin(2\theta)$ terms where necessary, and using eq.\ \ref{eq:wmp_and_aq}. This gives 
\begin{equation}
\overline{\gamma_+} = \frac{\gamma\sbr{+}\spr{m} + \gamma\sbr{+}\spr{q} \cos(2\theta) + \wpm\epsilon\sbr{+}\spr{int} + (1+\wpm)a\sbr{q}\cos(2\theta)\epsilon\sbr{+}\spr{int}}{(1 + \wpm)[1+a\sbr{q}\cos(2\theta)]}
\end{equation}
To simplify this expression, we note that, while $\wpm$ can be quite large, $a\sbr{q}$ is small and so Taylor expanding the denominator, keeping terms to first order in $a\sbr{q}$ and separating these into into monopole and quadrupole terms, we obtain
\begin{equation}
\overline{\gamma}\sbr{+}\spr{m} = \frac{\gamma\sbr{+}\spr{m} + \wpm\epsilon\sbr{+}\spr{int}}{1+\wpm}
\end{equation}
for the monopole, and 
\begin{equation}
 \overline{\gamma}\spr{q}\sbr{+} = \frac{\gamma\sbr{+}\spr{q} + a\sbr{q} (\epsilon\sbr{+}\spr{int} - \gamma\sbr{+}\spr{m}) }{1+\wpm}
 \end{equation}
for the quadrupole term with $\cos(2\theta)$ dependence. For the monopole, we see there the well-known multiplicative ``boost'' factor \citep{SheJohFri04}:
\begin{equation}
B(R)\equiv {1+\wpm(R)}\,
\end{equation} and an additional correction $\wpm\epsilon\sbr{+}\spr{int}$ due to the intrinsic alignments of the satellites. Because the satellites are radially aligned, 
$\epsilon\sbr{+}\spr{int} < 0$, this effect reduces the observed tangential shear.

Our results are expressed in terms of $\Delta \Sigma$, not shear, so multiplying all terms by $\invEcrit$, and defining $\Delta \Sigma\sbr{int} \equiv \invEcrit\epsilon\sbr{+}\spr{int}$, 
these become
\begin{equation}
    \Delta\overline{\Sigma\sbr{m}} = B^{-1}\left[\Delta\Sigma\sbr{m} + \wpm\, \Delta\Sigma\sbr{int} \right]
\label{eq:correct_mono}
\end{equation}
\begin{equation}
\ecrit\overline{\gamma}_+\spr{q} = B^{-1}\left[ \ecrit\gamma_+\spr{q} +a\sbr{q}(\Delta\Sigma\sbr{int} - \Delta\Sigma\sbr{m}) \right]
\label{eq:gamma_plus_corr}
\end{equation}
\begin{equation}
\ecrit\overline{\gamma}_{\times}\spr{q} = B^{-1}\,\ecrit\gamma_{\times}\spr{q}
\label{eq:gamma_cross_corr}
\end{equation}

In summary, for the quadrupole given by eq.\ \ref{eq:gamma_plus_corr}, apart from the multiplicative boost factor there are two new terms: the radial intrinsic alignment  $a\sbr{q}\Delta \Sigma\sbr{int}$ and what we refer to as the anisotropic boost factor $-a\sbr{q}\Delta\Sigma\sbr{m}$. In practice, both these corrections have the same sign, but the latter dominates over the former for $R \gtrsim 100$ kpc/$h$ for our lenses.

\subsection{Halo ellipticity Estimators}
\label{sec:quad_estimators}
Here we consider two different estimators of the halo ellipticity that differ in how the ellipticity is measured. The purpose of these estimators is to measure the quadrupole signal, which is proportional to the halo ellipticity, while cancelling any contribution from the monopole shear. 

The first of the quadrupole estimators used in this paper are from \cite{ClaJai16}, and will be referred to as the CJ estimators. These estimators measure the halo ellipticity directly, independently of the ellipticity of the galaxy. Moreover, they nullify the purely tangential monopole lensing signal. They provide 4 statistically independent measurements of halo ellipticity that we combine to calculate the mean halo ellipticity. A potential bias in the halo ellipticity arises if some effect aligns the major axis of the BCG in the same sense as the background sources. This might occur due to cosmic shear from foreground large-scale structure that is closer than the BCG. This effect may shear both the BCG and the background sources in the same sense. Alternatively, an uncorrected PSF anisotropy may also affect both the BCG and the background sources.  However, this systematic only affects the $\Delta\Sigma_1^{(+/-)}$ estimators and not the $\Delta\Sigma_2^{(+/-)}$ estimators \citep{ClaJai16}.

The second set of estimators assume that the halo ellipticity is related to the ellipticity of the lens galaxy stellar light,  $e_{\rm g}$. These estimators were first introduced by \cite{HoeYeeGla04}, further developed by \cite{Mandelbaum06}, and will be referred to here as Hoekstra-Mandelbaum (HM) estimators. The $f\Delta\Sigma$ and $f_{45}\Delta\Sigma$ estimators will both be affected by a systematic shear, as described above. However, to a good approximation, the two estimators will both be equally affected by this spurious shear  \citep{Mandelbaum06}, and therefore, 
by subtraction, we can measure the uncontaminated value $(f-f_{45})\Delta\Sigma$. \cite{Schrabback15} tested this with ray-tracing simulations and showed that the cancellation of systematics works well, at least for lenses at low redshift and for lens-source pairs at low projected galactocentric radii (see their fig. 6). These estimators are used to calculate the aligned ellipticity ratio $f\sbr{h} \sim e\sbr{h}/e\sbr{g}$.

Because the two sets of halo ellipticity estimators weight and combine the source shapes in different ways, they are statistically correlated, but not perfectly (nor of course are they statistically independent). We consider both estimators in this paper.

\subsubsection{Clampitt-Jain Estimators}

\begin{figure*}
\includegraphics[width=\linewidth]{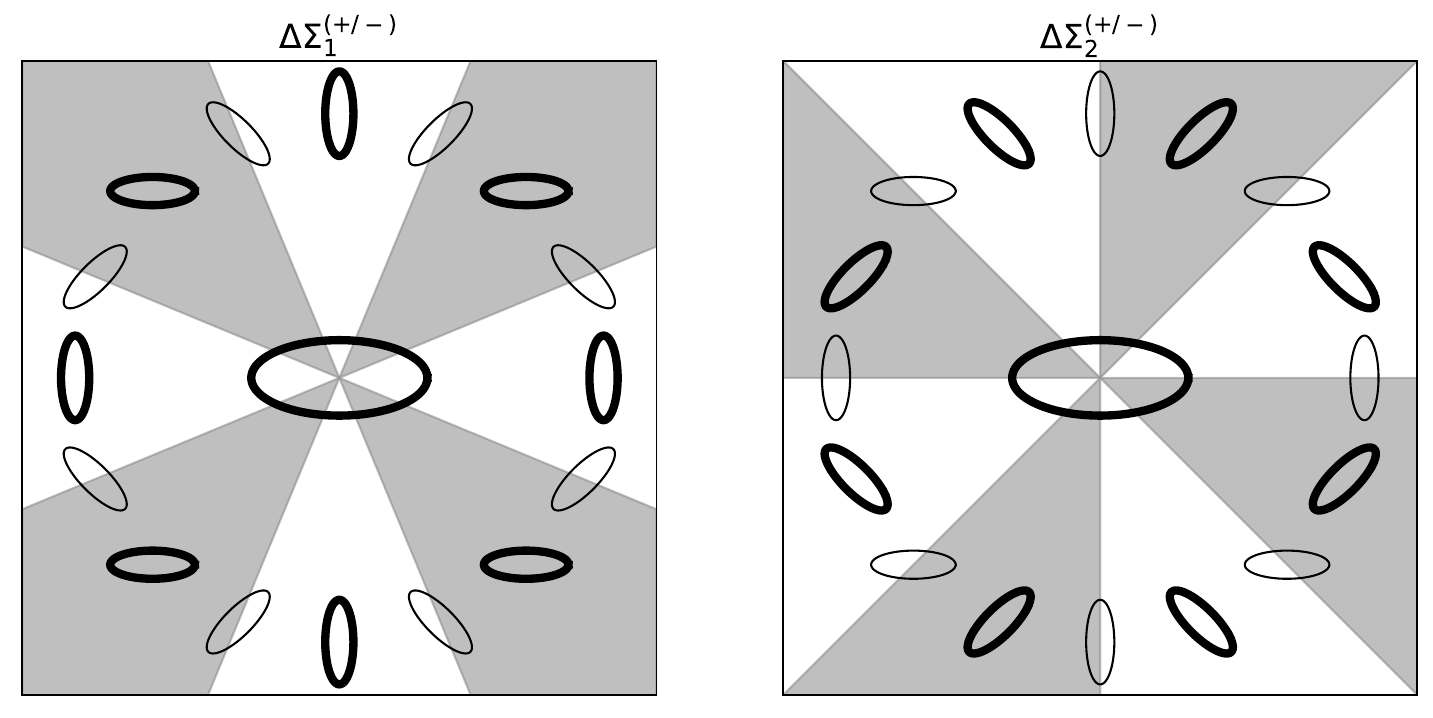}
\caption{A visualization of the 2D quadrupole shear pattern, and the regions covered by the CJ estimators. In both panels, the central ellipse reprsents the lens galaxy, while the smaller surrounding ellipses represent sources. These sources are experiencing a purely quadrupole shear. The left panel demonstrates the angular regions covered by the $\Delta\Sigma_1^{(+/-)}$ estimators. Sources in the lighter region are covered by $\Delta\Sigma_1^{-}$, while those in the darker region are covered by $\Delta\Sigma_1^{+}$. Sources which experience no $\gamma_1$ component have been drawn with a reduced line weight. The right panel demonstrates the angular regions covered by the $\Delta\Sigma_2^{(+/-)}$ estimators. Sources in the lighter region are covered by $\Delta\Sigma_2^{-}$, while those in the darker region are covered by $\Delta\Sigma_2^{+}$. Sources which experience no $\gamma_2$ component have been drawn with a reduced line weight.}
\label{fig:CJ_example}
\end{figure*}

The first set of estimators are based on those used by \cite{ClaJai16}. These estimators use the tangential and cross shear components of the quadrupole, $\ecrit\gamma_+$ and $\ecrit\gamma_{\times}$ \citep{Adhikari2015}. These are transformed into a coordinate system aligned with the major axis of the lens.
\begin{equation}
\ecrit\gamma_1 = -\ecrit\gamma\sbr{+}\cos2\theta + \ecrit\gamma_{\times}\sin2\theta
\label{eq:sigma_gamma1}
\end{equation}
\begin{equation}
\ecrit\gamma_2 = -\ecrit\gamma\sbr{+}\sin2\theta - \ecrit\gamma_{\times}\cos2\theta
\label{eq:sigma_gamma2}
\end{equation}
where
\begin{multline}
\ecrit\gamma_1(R) = (e/4)[(2I_1(R) - \Sigma\sbr{m}(R)\eta(R))\cos4\theta \\
+ 2I_2(R) - \Sigma\sbr{m}(R)\eta(R)]
\label{eq:sigma_gamma1}
\end{multline}
\begin{equation}
\ecrit\gamma_2(R) = (e/4)[2I_1(R) - \Sigma\sbr{m}(R)\eta(R)]\sin4\theta
\end{equation}

We correct for contributions from satellite galaxies by inserting equations (\ref{eq:gamma_plus_corr}) and (\ref{eq:gamma_cross_corr}) into equations (\ref{eq:sigma_gamma1}) and (\ref{eq:sigma_gamma2}).
\begin{equation}
\ecrit\overline{\gamma}_1=B^{-1}\left[\ecrit\gamma_1-\frac{a\sbr{q}}{2}(\Delta\Sigma\sbr{int}-\Delta\Sigma\sbr{m})(1+\cos4\theta)\right]
\label{eq:sigma_gamma1_corr}
\end{equation}
\begin{equation}
\ecrit\overline{\gamma}_2=B^{-1}\left[\ecrit\gamma_2 - \frac{a\sbr{q}}{2} (\Delta\Sigma\sbr{int}-\Delta\Sigma\sbr{m})\sin4\theta\right]
\label{eq:sigma_gamma2_corr}
\end{equation}
where $\ecrit\gamma_1$ and $\ecrit\gamma_2$ are equations (9) and (11) from \cite{ClaJai16}.

We use 4 estimators, divided into two pairs $\Delta\Sigma_1^{(+/-)}$ and $\Delta\Sigma_2^{(+/-)}$. The source galaxies will experience a quadrupole component in their shear, which depends on $e$. These estimators are designed to measure this quadrupole component in regions where $\gamma_1$ (or $\gamma_2$) have the same sign. A visualization of the CJ estimators, as well as the 2D quadrupole shear pattern, is presented in Fig.~\ref{fig:CJ_example}. In both panels, the central ellipse reprsents the lens galaxy, while the smaller surrounding ellipses represent sources. These sources are experiencing a purely quadrupole shear. In the left panel, sources in the lighter region are counted as part of the $\Delta\Sigma_1^{-}$ estimator. In this region, sources will experience a purely negative $\gamma_1$ component of quadrupole shear. Darker regions, where sources experience a positive $\gamma_1$ quadrupole shear, are measured with the $\Delta\Sigma_1^{+}$ estimator. Sources which experience no $\gamma_1$ component have been drawn with a reduced line weight. The right panel demonstrates the angular regions covered by the $\Delta\Sigma_2^{(+/-)}$ estimators. Sources in the lighter region are covered by $\Delta\Sigma_2^{-}$, experiencing a negative $\gamma_2$ quadrupole shear. Those in the darker region are covered by $\Delta\Sigma_2^{+}$, and experience a positive $\gamma_2$ quadrupole shear. Sources which experience no $\gamma_2$ component have been drawn with a reduced line weight.

The first pair of estimators, $\Delta\Sigma_1^{(+)}$ and $\Delta\Sigma_1^{(-)}$, depend on the $\gamma_1$ component.
\begin{equation}
\Delta\Sigma_1^{(-)}(R) = \frac{4}{\pi}\int_{-\pi/8}^{\pi/8} \ecrit\overline{\gamma}_1(R,\theta) d\theta
\label{eq:DS1plus} + \mbox{3 rotations by $\pi/2$}
\end{equation}
\begin{equation}
\Delta\Sigma_1^{(+)}(R) = \frac{4}{\pi}\int_{\pi/8}^{3\pi/8} \ecrit\overline{\gamma}_1(R,\theta) d\theta  + \mbox{3 rotations by $\pi/2$}
\end{equation}
where each of the bounds have 3 additional $\pi/2$ rotations so that they cover a \mbox{\Large$+$}  or  \mbox{\Large$\times$} shape on the sky.
The final pair of estimators, $\Delta\Sigma_2^{(+)}$ and $\Delta\Sigma_2^{(-)}$, depend on the $\gamma_2$ component.
\begin{equation}
\Delta\Sigma_2^{(-)}(R) = \frac{4}{\pi}\int_0^{\pi/4} \ecrit\overline{\gamma}_2(R,\theta) d\theta  + \mbox{3 rotations by $\pi/2$}
\end{equation}
\begin{equation}
\Delta\Sigma_2^{(+)}(R) = \frac{4}{\pi}\int_{\pi/4}^{\pi/2} \ecrit\overline{\gamma}_2(R,\theta) d\theta  + \mbox{3 rotations by $\pi/2$}
\label{eq:DS2minus}
\end{equation}

In practice, we measure these from the data with a weighted average of the source ellipticities
\begin{equation}
    \Delta\Sigma_k^{(s)} = \frac{\sum w\sbr{s}\epsilon_{k,\textrm{s}}\Sigma\sbr{crit,l} W\sbr{l}}{\sum w\sbr{s} W\sbr{l}}\,,
\label{eq:DS12data}
\end{equation}
where $k = {1,2}$ refers to which ellipticity component ($\epsilon_k$) is used in the weighted average. The angular range of the weighted average changes depending on $k$ and $s$.
\begin{equation}
\begin{split}
    k=1,s=- &: \quad -\pi/8 \leq \theta < \pi/8 \\
    k=1,s=+ &: \hspace{8pt}\quad \pi/8 \leq \theta < 3\pi/8 \\
    k=2,s=- &: \hspace{19pt}\quad 0 \leq \theta < \pi/4 \\
    k=2,s=+ &: \hspace{8pt}\quad \pi/4 \leq \theta < \pi/2
\end{split}
\end{equation}
with each of these having 3 other $\pi/2$ rotations so that they cover  a \mbox{\Large$+$}  or  \mbox{\Large$\times$} shape on the sky (see Figure \ref{fig:CJ_example}).

To obtain the halo ellipticity we compare to the measurements from equation~(\ref{eq:DS12data}) to our predictions from the model, which are calculated by inserting equations (\ref{eq:sigma_gamma1_corr}) and (\ref{eq:sigma_gamma2_corr}) into equations (\ref{eq:DS1plus}) -- (\ref{eq:DS2minus}), then evaluating the integrals. We can fit each of the resulting equations to find $e$.

\subsubsection{Hoekstra-Mandelbaum Estimators}
The HM estimators assume that the halo ellipticity depends on the lens galaxy stellar ellipticity, $e_{\rm g}$. Their model for the anisotropic mass distribution is slightly different from the one present in equation (\ref{eq:aniso_model}).
\begin{equation}
\Delta\Sigma(R, \theta) = \Delta\Sigma\sbr{m}(R)\left[ 1+2f(R)e\sbr{g} \cos(2\theta) \right]\,.
\end{equation}
where $f$ is a factor that relates the ellipticity of the galaxy light ($e_g$) to the ellipicity of the halo (e). This is similar to the model described by equation (\ref{eq:aniso_model}), with slightly different notation, where $f(R) = -\frac{e}{e\sbr{g}}\eta(R)$. Note that this model differs by a factor of 2 from the model used by \cite{Schrabback15}, who adopt a different definition of lens galaxy ellipticity.

This yields two estimators, $f\Delta\Sigma$ and $f_{45}\Delta\Sigma$, that depend on the tangential and cross ellipticities $(\epsilon\sbr{t}$ and $\epsilon\sbr{\times})$.
\begin{equation}
f(R)\Delta\Sigma\sbr{m}(R) = \frac{\sum_i\epsilon_{{\rm +},i}\langle\ecrit^{-1}(z\sbr{l})\rangle^{-1} w_i W_l e_{{\rm g},i} \cos(2\theta_i)}{2 \sum_i w_i W_l e_{{\rm g},i}^2\cos^2(2\theta_i)}
\label{eq:MAN_est1}
\end{equation}
\begin{equation}
f_{45}(R)\Delta\Sigma\sbr{m}(R) = \frac{\sum_i\epsilon_{{\rm \times},i}\langle\ecrit^{-1}(z\sbr{l})\rangle^{-1} w_i W_l e_{{\rm g},i} \sin(2\theta_i)}{2 \sum_i w_i W_l e_{{\rm g},i}^2\sin^2(2\theta_i)}
\label{eq:MAN_est2}
\end{equation}
where $W\sbr{l}=\langle\ecrit^{-1}(z\sbr{l})\rangle^2$ and
\begin{equation}
\epsilon\sbr{+} = -\epsilon_1\cos 2\theta - \epsilon_2 \sin 2\theta \,,
\end{equation}
\begin{equation}
\epsilon_{\times} = +\epsilon_1 \sin 2\theta - \epsilon_2 \cos 2\theta \,.
\end{equation}
These estimators measure the anisotropic component of the shear, or the quadrupole. The $\cos(2\theta)$ weighting in equation~(\ref{eq:MAN_est1}) will apply a positive weighting along the major axis of the lens, where we expect $\epsilon\sbr{+}$ to be positive. However, it will apply a negative weighting along the minor axis, where we expect $\epsilon\sbr{+}$ to be negative. If the lenses and sources are aligned due to systematic effects, rather than the lens shear, these will effect this estimator. We define a second estimator, equation~(\ref{eq:MAN_est2}), which depends on the cross shear, $\epsilon\sbr{\times}$. This estimator will experience the same systematic effects experienced by the first estimator, allowing us to cancel these contributions. Therefore, an advantage of the HM estimator $(f-f_{45})\Delta\Sigma\sbr{m}(R)$ is that it cancels systematics.

It is worth noting that the sign convention for the shears are different than for the CJ estimators. For example, a source at an angle of $\theta = 0$ from the major axis experiencing a purely tangential shear would have a negative $\epsilon_1$ component. However, this would be a positive tangential shear, $\epsilon\sbr{+}$.

To model the expected signal from the HM estimators, we  replace the weighted averages of $\epsilon$ in equations (\ref{eq:MAN_est1}) and (\ref{eq:MAN_est2}) with the models for $\overline{\gamma}\sbr{+}$ and $\overline{\gamma}\sbr{\times}$ and integrate $\theta$ from 0 to $2\pi$, yielding
\begin{equation}
f(R)\Delta\Sigma\sbr{m}(R) = \frac{1}{2\pi \overline{e}\sbr{g}} \int_0^{2\pi}\ecrit\overline{\gamma}\sbr{+}\spr{q}(R) \cos^2(2\theta)
\label{eq:fsigma_integral}
\end{equation}
\begin{equation}
f_{45}(R)\Delta\Sigma\sbr{m}(R) = \frac{1}{2\pi \overline{e}\sbr{g}} \int_0^{2\pi}\ecrit\overline{\gamma}\sbr{\times}\spr{q}(R) \sin^2(2\theta)
\label{eq:f45sigma_integral}
\end{equation}
where $\overline{e}\sbr{g}$ is the average ellipticity of the lens galaxy light. We correct for satellite contamination by using $\ecrit\overline{\gamma}\sbr{+}\spr{q}$ and $\ecrit\overline{\gamma}\sbr{\times}\spr{q}$ from equations (\ref{eq:gamma_plus_corr}) and (\ref{eq:gamma_cross_corr}). We continue to use the $a\sbr{q}$ and $\wpm$ as described in Section \ref{sec:satellites} and Fig. \ref{fig:angular_bins}. Dividing the lens sample into different $e_g$ bins did not significantly affect the fits of $a\sbr{q}$ and $\wpm$. So we treat the fits as valid for all lenses, regardless of $e_g$.

After inserting  eqs. (\ref{eq:gamma_plus_corr}) and (\ref{eq:gamma_cross_corr})  into equations (\ref{eq:fsigma_integral}) and (\ref{eq:f45sigma_integral}), and performing the integrals, we find
\begin{equation}
f\Delta\Sigma\sbr{m} = B^{-1}\left[\frac{f_h}{4}(\Sigma\sbr{m}\eta - I_1 -I_2)
+\frac{a\sbr{q}}{2\overline{e}_g}(\Delta\Sigma\sbr{int} - \Delta\Sigma\sbr{m})\right]
\end{equation}
\begin{equation}
f_{45}\Delta\Sigma\sbr{m} = \frac{f_h}{4}B^{-1} [-I_1+I_2]
\end{equation}
To reduce the effect of systematics, we subtract $f_{45}\Delta\Sigma$ from $f\Delta\Sigma$ giving
\begin{equation}
(f-f_{45})\Delta\Sigma\sbr{m} =  B^{-1}\left[\frac{f_h}{4}(\Sigma\sbr{m}\eta - 2I_2
+\frac{a\sbr{q}}{2\overline{e}_g}(\Delta\Sigma\spr{int} - \Delta\Sigma\sbr{m})\right]
\label{eq:f_subtract_model}
\end{equation}

In summary, we measure $f\Delta\Sigma\sbr{m}$ and $f_{45}\Delta\Sigma\sbr{m}$ in radial bins using the equations (\ref{eq:MAN_est1}) and (\ref{eq:MAN_est2}). Then we subtract these to obtain $(f-f_{45})\Delta\Sigma\sbr{m}$, which we  fit using equation (\ref{eq:f_subtract_model}) to obtain $f_h$.

\subsubsection{Misalignment between the position angles of the stellar light and the dark matter halo}
\label{sec:misalign}
We expect the light of the LRGs to be aligned with their dark matter haloes, making the major axis of the light a proxy for the major axis of the dark matter halo. The alignment, however, is not perfect and has been studied by a number of authors. 

If the probability density function of misalignment angles is $P(\theta\sbr{mis})$, then the aligned ellipticity $e\sbr{eff}$ will be reduced from the halo ellipticity by a factor \citep{ClaJai16}
\begin{equation}
    \frac{e\sbr{eff}}{e} = \langle\cos(2\theta\sbr{mis})\rangle = \int \cos(2\theta\sbr{mis}) P(\theta\sbr{mis}) d\theta\sbr{mis}\,.
    \label{eq:eeff}
\end{equation}
For example, if one assumes $P(\theta\sbr{mis})$ is a Gaussian distribution with width $\sigma\sbr{mis}$ in radians, then
\begin{equation}
\frac{e\sbr{eff}}{e} = \exp(-2\sigma\sbr{mis}^2)
\label{eq:misGauss}
\end{equation}

One way to assess the degree of misalignment is through hydrodynamical simulations of galaxy formation that  predict the distribution of misalignment angles, $P(\theta\sbr{mis})$. A number of authors have studied the misalignment angle between the projected stellar distribution and the projected total matter distribution in various halo mass bins. They find the misalignment angle tends to be smaller in more massive haloes. In Appendix \ref{sec:Vel15}, we compute relevant quantities for the case of \cite{VelCacSch15}. We return to the topic of misalignment in the discussion in Section~\ref{sec:discussion}.

\section{Results}
\label{sec:results}

\subsection{Weak Lensing Monopole}
Before fitting the elliptical model to the quadrupole shear, we require the monopole of the shear. The monopole of the tangential shear was measured in radial bins and scaled using $\ecrit$ for each lens galaxy. Then an NFW profile was fit with halo mass and concentration as free parameters. Only bins within 1 Mpc/$h$ from the centre of the lens were included in the fit, as beyond this radius effects from other haloes and surrounding structures become significant. A sample of 10,000,000 random lenses was created, and the lensing signal around these random lenses was measured. Before fitting, we subtract the results of the random lenses from the measured signal.

\begin{figure}
\includegraphics[width=\columnwidth]{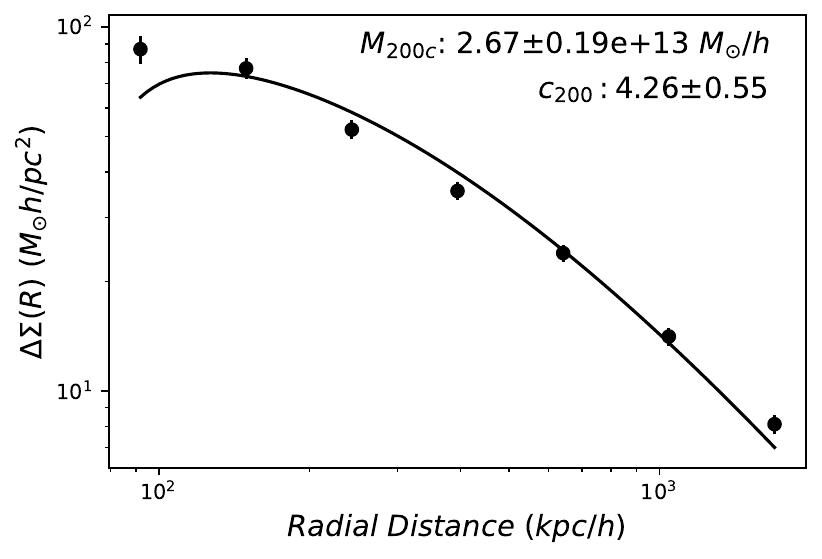}
\caption[Monopole Shear Results]{The monopole tangential shear of the DR7 LRGs with the best fitting NFW halo fit using  bins within 1 Mpc/$h$. The observed monopole shear was fit with the corrected model in equation~(\ref{eq:correct_mono}). The best-fitting halo mass and concentration are displayed in the plot.}
\label{fig:results_mono}
\end{figure}

The results of the stacking and the fit for the DR7 LRGs are shown in Fig.~\ref{fig:results_mono}. Both parameters are quite well constrained, and there is good agreement between the data within 1 Mpc/$h$ and the NFW model. We obtain a halo mass of $M_{200c} = (2.67 \pm 0.19) \times 10^{13}$ M$_{\odot}/h$ and a concentration of $c_{200c} = 4.26 \pm 0.55$.

The mass is consistent with that found by \cite{MandelbaumSeljakCool2006}, for which a lens-weighted average over their bright and faint DR4 LRG subsamples yields $(3.04 \pm 0.39) \times 10^{13}$ M$_{\odot}/h$,  after conversion to our mass definition. Similarly converted, their concentration is $2.8\pm0.4$, which is lower than our fit. We note, however, that while \cite{MandelbaumSeljakCool2006} used photometric redshifts for lenses and bright sources to reduce the contamination of satellite galaxies, this may not eliminate contamination entirely. Indeed, the fact that they have a boost factor $B > 1$ implies there is some contamination. They do not model the effects of radial intrinsic alignments or an \emph{anisotropic} boost factor. If we had ignored these corrections, we would have found a concentration of $3.0\pm0.3$, consistent with their results.

The concentration from the fit is in agreement with the concentration predicted by mass-concentration relations. For a halo of this mass, \cite{DufSchKay08} predict a concentration $c_{200c}\sim 4$, while \cite{DutMac14} predict $c_{200c}\sim 5$.

For the BOSS LRGs, we obtain a halo mass of $M_{200c} = (1.15 \pm 0.09) \times 10^{13}$ M$_{\odot}/h$ and a concentration of $c_{200c} = 2.97 \pm 0.48$.

\subsection{Weak Lensing Quadrupole}

\begin{figure*}
\includegraphics[width=\linewidth]{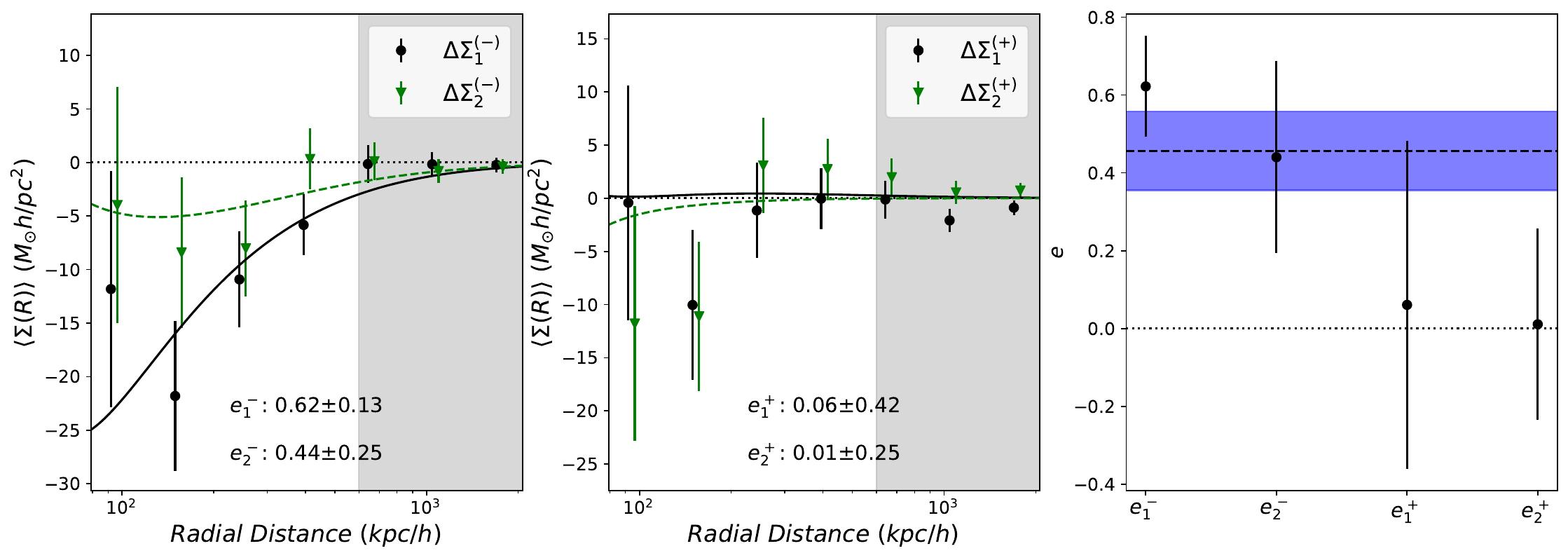}
\caption{Quadrupole shear around DR7 LRGs aligned with the major axis of the LRG light. Results obtained after correcting for systematics are shown. The left panel displays the negative CJ estimators, while the centre panel displays the positive CJ estimators. The first estimator is represented in black with circles and a solid line. The second estimator is represented in green with triangles and a dashed line. Points plotted in green have been shifted slightly to the right for clarity. The best fit of $e$ for each estimator is displayed in each panel. The right panel displays the halo ellipticity values from the 4 independent CJ estimators. The weighted average halo ellipticity $(e=0.46)$ is plotted as a dashed black line. The range of 1$\sigma$ uncertainty in the mean is shaded in blue $(\Delta e=0.10)$. No ellipticity is represented with a dotted black line at $e=0$.}
\label{fig:results}
\end{figure*}

\begin{figure*}
\includegraphics[width=0.7\linewidth]{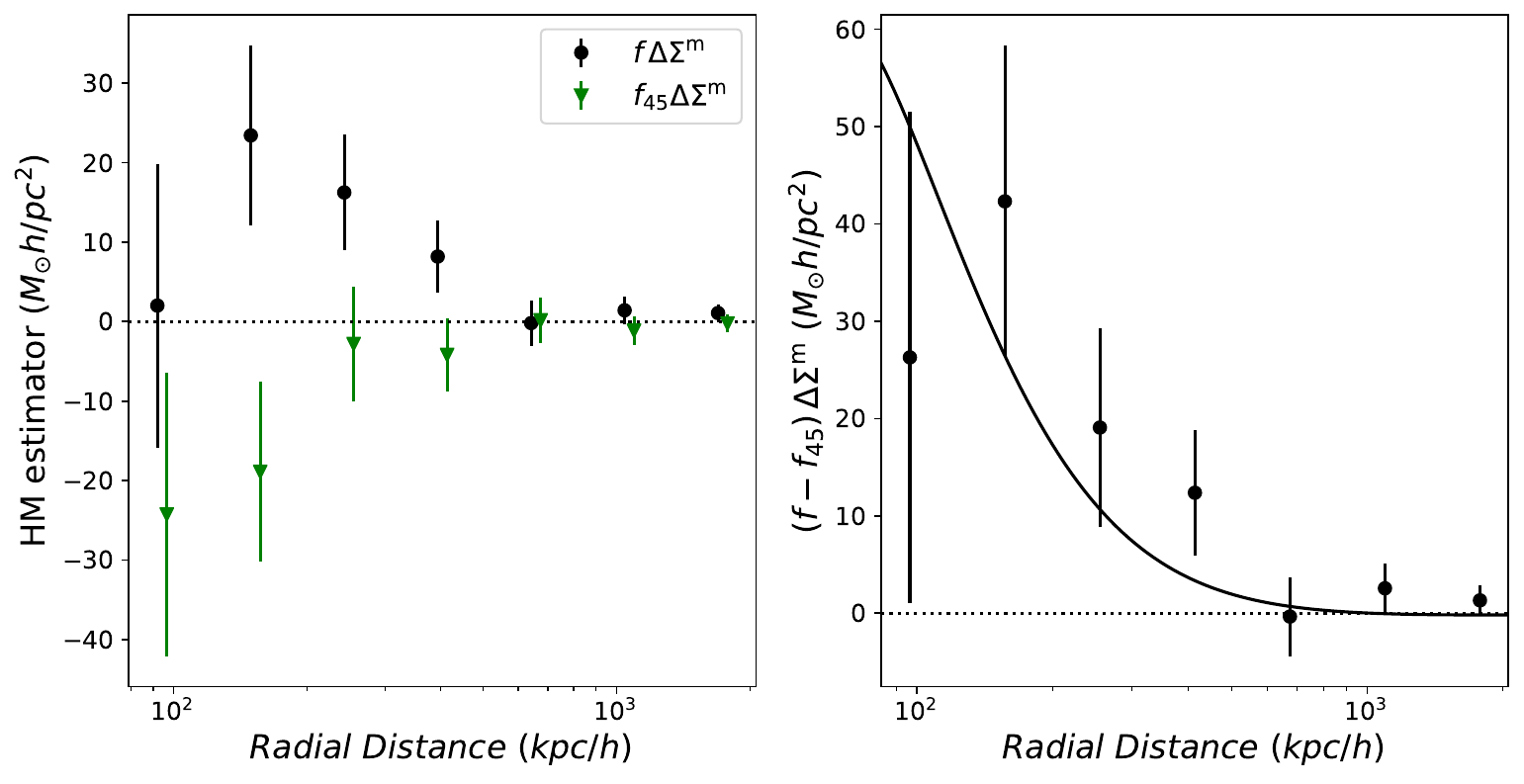}
\caption{The left panel displays the HM estimators from equations (\ref{eq:MAN_est1}) and (\ref{eq:MAN_est2}) for DR7 lenses. $f(R)\Delta\Sigma\sbr{m}(R)$ is represented by black circles, while $f_{45}(R)\Delta\Sigma\sbr{m}(R)$ is represented by green triangles. The right panel displays the difference $(f-f_{45})\Delta\Sigma\sbr{m}(R)$. The black line represents a fit with equation (\ref{eq:f_subtract_model}). From this fit we obtain $f_h = 2.2 \pm 0.6$.}
\label{fig:chi2_e}
\end{figure*}

Results for the CJ estimators applied to the DR7 LRGs are displayed in Fig.~\ref{fig:results}. A model based on the halo mass and concentration for the monopole but with free ellipticity $e$ was fit to each of the 4 quadrupole estimators independently. Only radial bins within 600 kpc/h were included in the fit, as surrounding structure contributes a significant amount of anisotropy at large radii. The region not included in the fit is shaded in gray. The best fit of $e$ for each CJ estimator is displayed in the appropriate panel.

We can quantify the significance of the agreement between these ellipticity values from the fits. A weighted average of the halo ellipticity is taken using the uncertainty in the value of $e$ from the fit ($w=\sigma_e^{-2}$). The 4 ellipticities from the CJ estimators are independent of each other, because they cover different angular ranges around the lens and different components of the ellipticity. From the 4 CJ estimators, the mean halo ellipticity is $e = 0.46 \pm 0.10$. If we assume these 4 values of ellipticity are fit with a constant $e$, we can calculate the $\chi^2$ of this hypothetical fit, which is $\chi^2 = 5.79$. This is dominated by the $\Delta \Sigma_2^-$ estimator which contributes 3.27. Thus the deviation of this points is at a level less than 2$\sigma$. We can then evaluate the cumulative distribution function for a $\chi^2$ with 3 degrees of freedom. We find a 12\% chance of obtaining a value of $\chi^2 = 5.79$ or higher, which is acceptable.

We show the constant model and the data in Fig.~\ref{fig:results}. When these estimators are fit with an ellipticity of $e=0$, the value raises to $\chi^2 = 26.19$, with a probability of 0.0008\%.

Results for the HM estimators are displayed in Fig.~\ref{fig:chi2_e}. For the DR7 LRGs, we find $f_h = 2.2 \pm 0.6$. These results are consistent given the mean ellipticity of the galaxy light for the lens sample, $e\sbr{g}=0.22$ predicts a halo ellipticity $e = 0.48\pm0.08$. This is a significant detection of a non-zero $f_h$.

\begin{figure*}
\includegraphics[width=\linewidth]{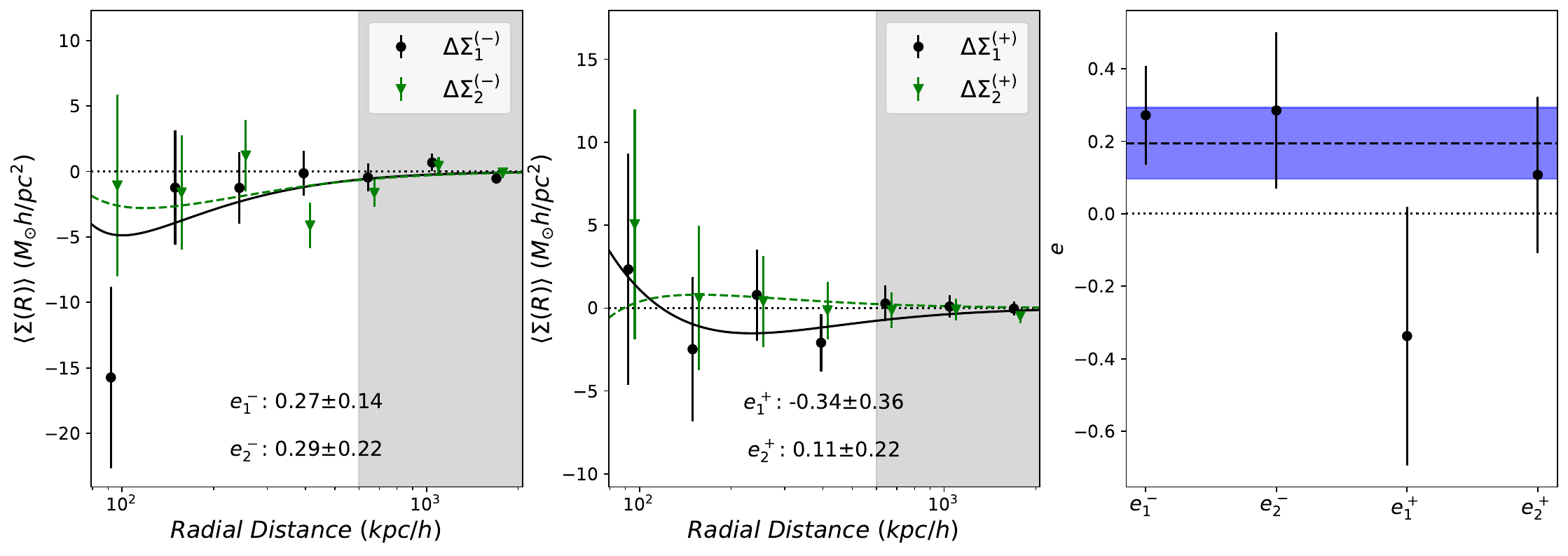}
\caption{CJ estimators applied to lenses from CMASS and LOWZ samples of BOSS. The left panel displays the negative CJ estimators, while the centre panel displays the positive CJ estimators. The first estimator is represented in black with circles and a solid line. The second estimator is represented in green with triangles and a dashed line. Points plotted in green have been shifted slightly to the right for clarity. The best fit of $e$ for each estimator is displayed in each panel. The right panel displays the halo ellipticity values from the 4 independent CJ estimators. The weighted average halo ellipticity $(e=0.20)$ is plotted as a dashed black line. The range of 1$\sigma$ uncertainty in the mean is shaded in blue $(\Delta e=0.10)$. No ellipticity is represented with a dotted black line at $e=0$.}
\label{fig:results_old}
\end{figure*}

\begin{figure*}
\includegraphics[width=0.7\linewidth]{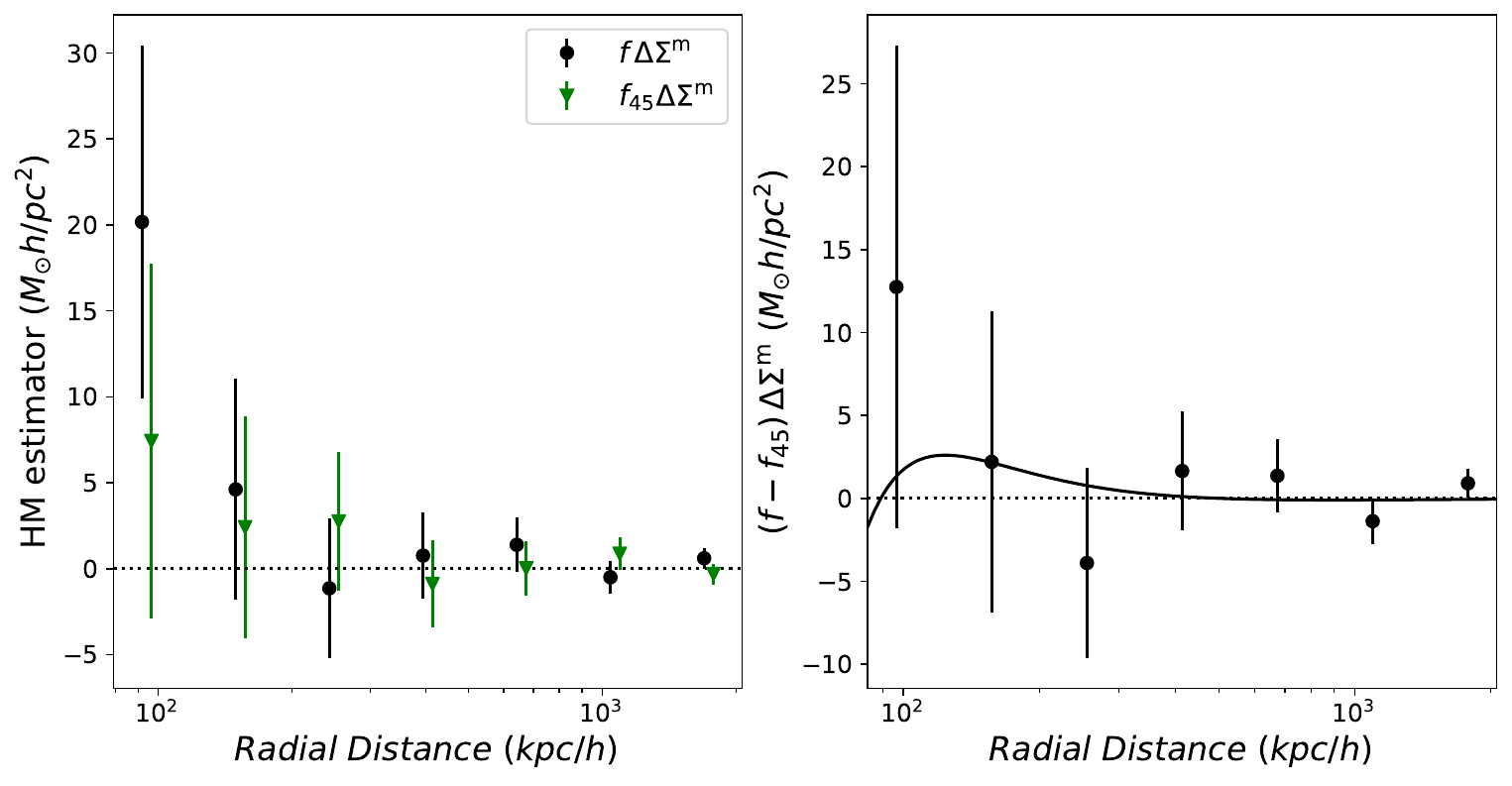}
\caption{The left panel displays the HM estimators for the the CMASS and LOWZ samples of BOSS. $f(R)\Delta\Sigma\sbr{m}(R)$ is represented by black circles, while $f_{45}(R)\Delta\Sigma\sbr{m}(R)$ is represented by green triangles. The right panel displays the difference $(f-f_{45})\Delta\Sigma\sbr{m}(R)$. From this fitting equation (\ref{eq:f_subtract_model}), we obtain $f_h = 0.7 \pm 0.7$.}
\label{fig:chi2_old}
\end{figure*}

Repeating the analysis for the BOSS LRGs, we obtain a mean halo ellipticity of $e = 0.20 \pm 0.10$. Using the HM estimators, we obtain $f_h=0.7 \pm 0.7$. The results for the CJ estimators are shown in Fig.~\ref{fig:results_old}, while the results for the HM estimators are displayed in Fig.~\ref{fig:chi2_old}. The mean ellipticity of the galaxy light for this lens sample is $e\sbr{g}=0.22$, so the HM fit predicts $e = 0.15\pm0.15$. Therefore, the values of $e$ and $f_h$ are consistent.

The BOSS haloes appear to be less elliptical than the DR7 LRG haloes, but in fact neither the difference in $e$ ($0.26\pm0.14$), nor in $f_h$ ($1.5\pm0.9$), is statistically significant. Note that the BOSS LRGs are slightly less massive than the SDSS DR7 LRGs, which may cause them to be less elliptical (see discussion in  Section~\ref{sec:discussion}). Moreover, the BOSS LRG sample is more distant $(z\sbr{median}=0.51)$ than the SDSS DR7 LRGs $(z\sbr{median}=0.34)$. At these redshifts there is loss of surface brightness, not only due to cosmological $(1+z)^4$ dimming but also because the CFHT $r$-band probes the rest-frame ultraviolet below the 4000 \AA\ break, where the flux from old, red galaxies is much suppressed. This may lead to less accurate measurements of the LRG ellipticity and of the major-axis position angle of the galaxy light, leading to greater misalignment when stacking. Also, for the above reasons, photometry may be more sensitive to the inner regions of the BOSS LRGs, whereas for the SDSS DR7 LRGs it may be more sensitive to the outer regions. In an elliptical galaxy with isophote twists, the latter may be better aligned with the DM halo. Finally, \citet{Schrabback15}, in their fig.~6., have shown that the HM estimators of $f\sbr{h}$ are biased low due to cosmic shear when the data extend to high galactocentric projected radii for lenses in the redshift range of the BOSS LRGs, whereas this bias is negligible for the lower redshift SDSS DR7 LRGs.

\section{Systematic tests: lens major axis position angles from SDSS}
\label{sec:systematics}
\begin{figure}
\includegraphics[width=\columnwidth]{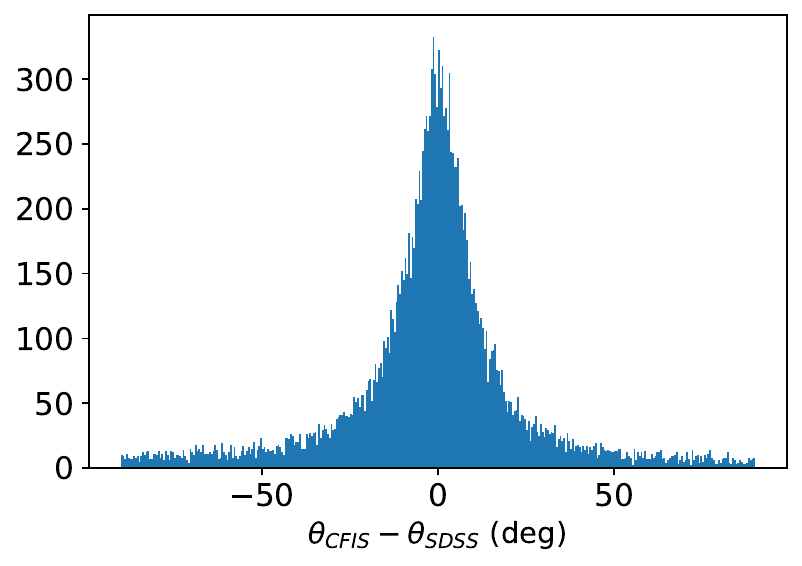}
\caption[LRG Angle Compare]{Comparison of the  position angles of the major axis of light in lens galaxies measured in UNIONS and in SDSS.}
\label{fig:angle_compare}
\end{figure}

When using lens position angles from UNIONS, the shape measurements of the sources and lenses are both derived from the same imaging. It is possible that issues with the shape measurements, for example inadequate PSF correction, could lead to a correlation between the lens and source shapes. This correlation could lead to an observed alignment that could affect our weak lensing results. To provide an alternate test, and an independent source of lens position angles, we can perform the quadrupole shear measurement using position angles from the SDSS database. First, all LRGs were selected from the DR16 release of SDSS. The resulting list of LRGs were then matched in equatorial coordinates to our list of LRGs that overlap with UNIONS. In SDSS, these LRGs were fit with both an exponential and a de Vaucouleurs profile. The angle from the fit with the highest likelihood was chosen. It is worth noting that the SDSS imaging is considerably shallower than UNIONS, so we might expect their position angles to be less accurate than the position angles from UNIONS. A histogram of the differences in position angle from UNIONS and SDSS is displayed in Fig.~\ref{fig:angle_compare}. Most position angles are similar, however, there is a significant number of LRGs that have substantially different major-axis position angles in SDSS photometry versus UNIONS photometry. The semi-interquartile range of the difference is $31.4^{\circ}$ and the standard deviation is $33.6^{\circ}$.

\begin{figure*}
\includegraphics[width=\linewidth]{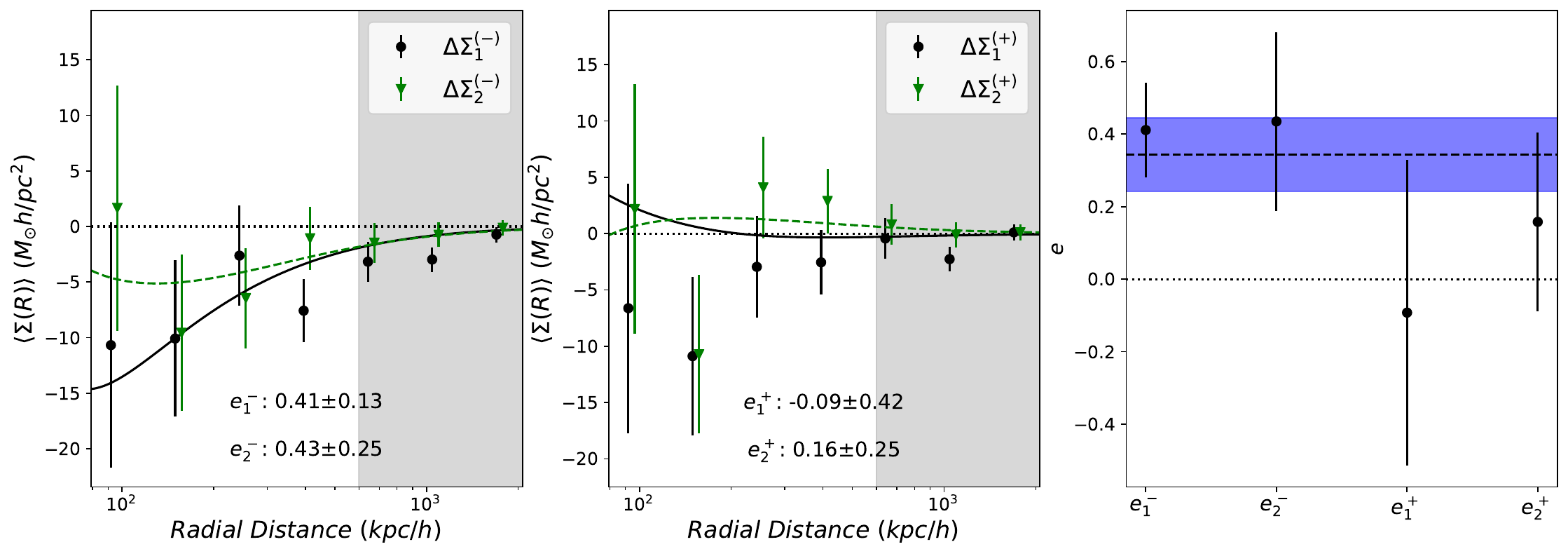}
\caption{CJ estimators applied to the DR7 LRGs. The lens major axis position angle from SDSS was used, as opposed to the position angle from UNIONS. The left panel displays the negative CJ estimators, while the centre panel displays the positive CJ estimators. The first estimator is represented in black with circles and a solid line. The second estimator is represented in green with triangles and a dashed line. Points plotted in green have been shifted slightly to the right for clarity. The best fit of $e$ for each estimator is displayed in each panel. The right panel displays the halo ellipticity values from the 4 independent CJ estimators. The weighted average halo ellipticity $(e=0.34)$ is plotted as a dashed black line. The range of 1$\sigma$ uncertainty in the mean is shaded in blue $(\Delta e=0.10)$. No ellipticity is represented with a dotted black line at $e=0$.}
\label{fig:results_sdss}
\end{figure*}

The process of calculating the quadrupole shear was repeated using the major axis light position angles from SDSS. Results for the quadrupole shear using the SDSS lens position angle are displayed in Fig.~\ref{fig:results_sdss}. We can repeat the process used for the UNIONS major axis position angles and calculate the mean halo ellipticity and $\chi^2$ of the fit. The mean halo ellipticity is $e~= 0.34 \pm 0.10$ and $\chi^2_{\nu} = 0.68$.  This is larger than the ellipticity obtained by \cite{ClaJai16}, $e=0.24\pm0.06$, who used SDSS photometery to study the same set of LRGs. However, they applied no correction for intrinsic radial alignments nor for the anisotropic boost. If we neglect these corrections, we find $e=0.25\pm0.09$, in good agreement with the results of \cite{ClaJai16}. From the HM estimators we obtain $f_h = 1.9 \pm 0.6$. The mean ellipticity from the SDSS position angles is lower than for the UNIONS position angles. This is not surprising: as mentioned before, the imaging from SDSS is shallower than it is for UNIONS, so we expect the position angles to be less accurate. This could lead to a higher degree of misalignment which would yield a rounder stacked shear.

To assess this effect, we note that the misalignment between the UNIONS major axis and the SDSS major axis will obey some probability distribution, $P(\theta\sbr{mis})$, which describes how likely the two axes will be separated by a given misalignment angle, as shown in Fig.~\ref{fig:angle_compare}. In Sec.~\ref{sec:misalign} we discussed the effect of a misalignment between the stellar light and the DM halo. However, observational errors due to signal-to-noise when measuring the major axis of the stellar light will also contribute to the misalignment. Moreover, differences may also arise because  elliptical galaxies have isophote twists, and so deeper photometry may probe outer regions of the galaxies which may have a different ellipticity and position angle than the inner regions. 

If the SDSS light angles are misaligned, they will yield a lower $e_{\textrm{eff}}$ than the ellipticity we measure from the UNIONS light angles, $e$, as given by equation~(\ref{eq:misGauss}). In Fig.~\ref{fig:angle_compare}, we found a standard deviation between the UNIONS and SDSS angles of $\sigma=33.6^{\circ}$. If we attribute all of the misalignment in the lens positions to errors in the SDSS photometry, using the standard deviation from Fig.~\ref{fig:angle_compare}, we should obtain an effective ellipticity of $e_{\textrm{eff}}=0.230$ for SDSS. This is somewhat lower than the observed value. In reality, there will be some measurement uncertainty in both the UNIONS and SDSS major axis position angles, although we expect the latter to be larger due to the shallower depth of the photometry.

\begin{figure*}
\includegraphics[width=0.7\linewidth]{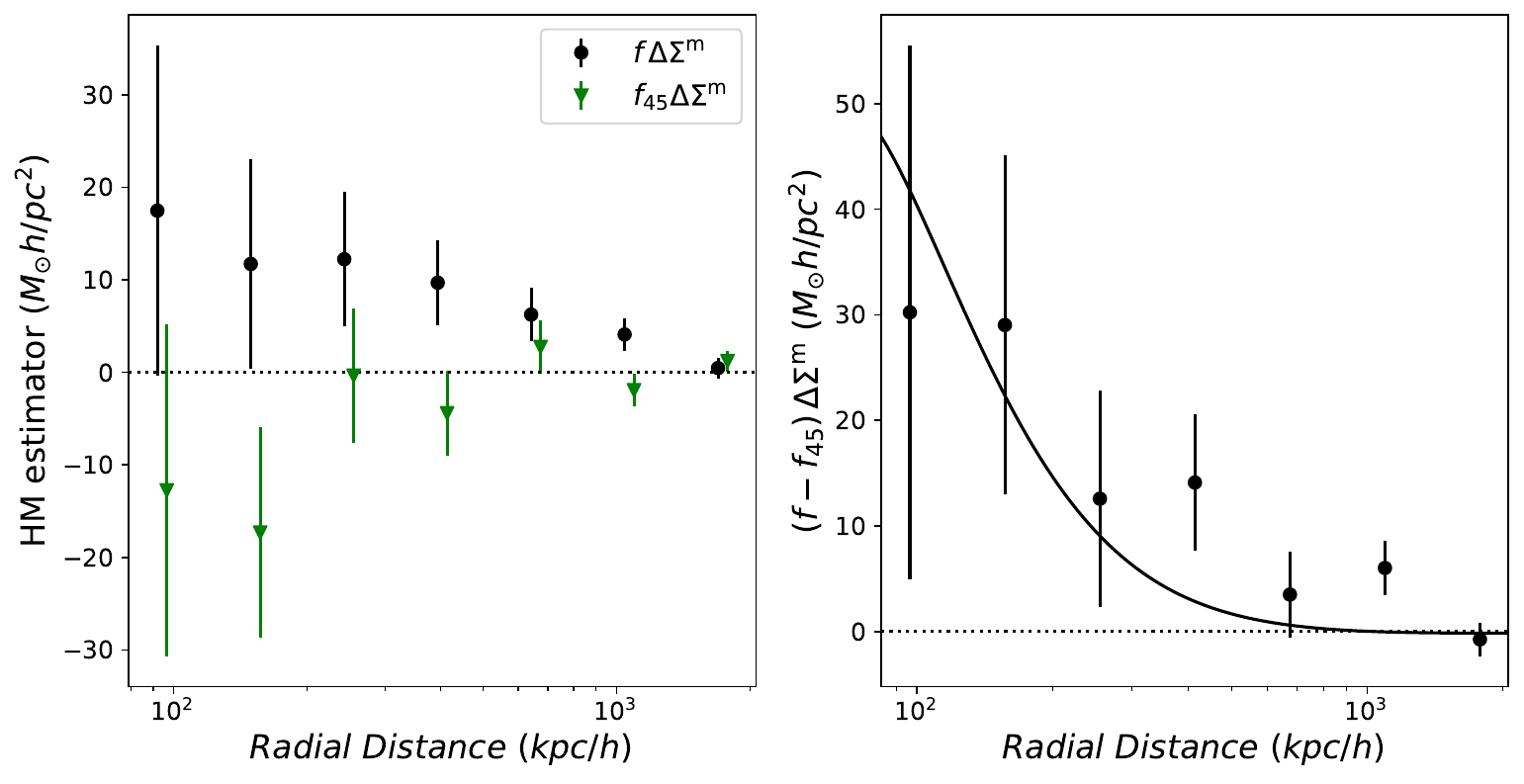}
\caption{The left panel displays the HM estimators using DR7 lenses with major axis position angles from SDSS. $f(R)\Delta\Sigma\sbr{m}(R)$ is represented by black circles, while $f_{45}(R)\Delta\Sigma\sbr{m}(R)$ is represented by green triangles. The right panel displays the difference $(f-f_{45})(R)\Delta\Sigma\sbr{m}(R)$. From this fitting equation (\ref{eq:f_subtract_model}), we obtain $f_h = 1.9 \pm 0.6$.}
\label{fig:chi2_e_sdss}
\end{figure*}

\section{Discussion}
\label{sec:discussion}

\begin{figure*}
\includegraphics[width=0.8\linewidth]{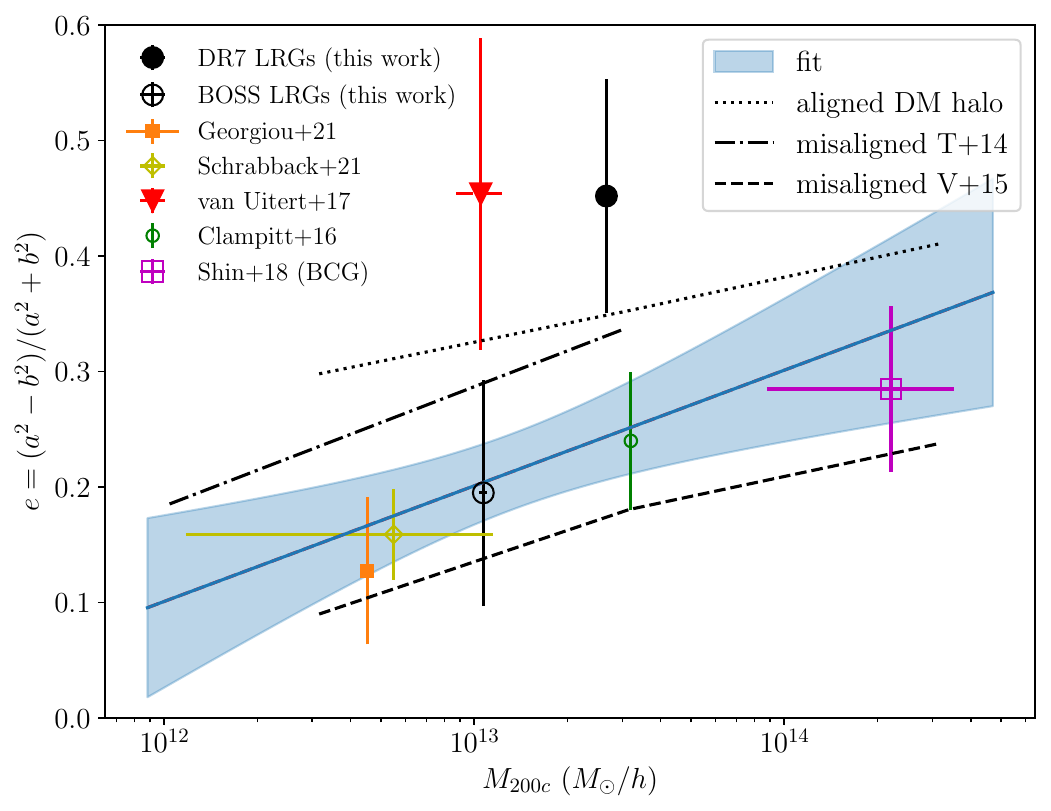}
\caption{Comparison of constraints on halo ellipticity $e$ as a function of halo mass for red galaxies, groups, and clusters. Where present, horizontal error bars represent approximate lens mass ranges for each study. We show the results from this paper and the papers cited in the text. The blue shaded band shows the fit to all data points, as described in the text. The dotted line indicates the expected trend from N-body simulations assuming DM halos are perfectly aligned with the stellar light whereas the dot-dash and dashed lines show the predictions using the misalignment models of 
\protect\cite{Tenneti14} 
and 
\protect\cite{VelCacSch15},
respectively.}
\label{fig:compare_e}
\end{figure*}

Fig.~\ref{fig:compare_e} compares our results for the halo ellipticity $e$ with others from the literature that are based on aligning the weak lensing signal with the major axis of red galaxies.  Studies included are: the red lens galaxies from \cite{Georgiou21} and \cite{Schrabback21}; the group-centrals from \cite{VanHoeJoa17}, where we use their fit over the range of $28-525$ kpc$/h$, which is similar to our fitted range; the study of DR7 LRGs by \cite{ClaJai16}, and the cluster study of \cite{ShiClaJai18}, where we use their result for alignment between the BCG and the cluster halo. We have converted results to use our conventions for ellipticity, $e$, and mass, $M_{200c}$. We make no attempt, however, to correct for different treatment of satellite contamination or different ways in which the major axis position angle is measured (see discussion in Section \ref{sec:systematics}). Horizontal error bars, where present, represent approximate lens mass ranges for each study.

To test whether the apparent trend of increasing ellipticity with halo mass is statistically significant, we fit a straight line to $e$ as a function of $M_{200c}$ using the data plotted in Fig.~\ref{fig:compare_e}. The resulting fit has a $\chi^2$ of 8.8, which, for 5 degrees of freedom, is acceptable ($p = 0.11)$. The best fit parameters are 
\begin{equation}
    e = (0.20\pm0.03) + (0.10\pm0.06)
     \log_{10} \left( \frac{M_{200c}}{10^{13} \Msun/h}\right).
     \label{eq:eslope}
\end{equation}
The increase in ellipticity with halo mass is not statistically significant. 

\begin{figure}
\includegraphics[width=\columnwidth]{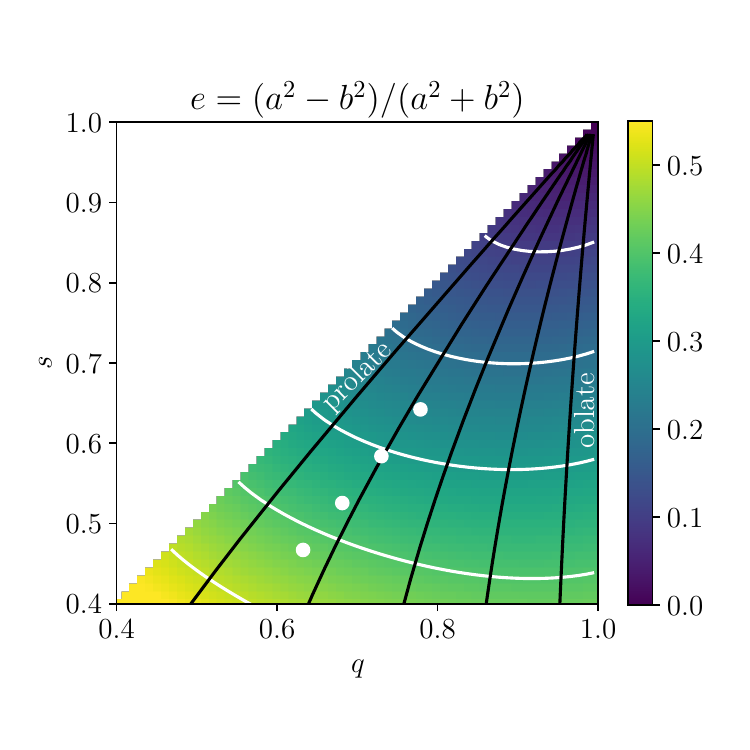}
\caption{The projected 2D ellipticity, $e$, for triaxial DM halos with axis ratios 1:$q$:$s$, as a function of $q$ and $s$, averaged over random projection angles is shown by the colour bar and the white contours (steps of 0.1).  Round halos are at the top right corner, prolate halos along the diagonal axis and oblate halos along the right hand axis. Black lines are constant triaxiality $T  = (1-q^2)/(1-s^2) = 0.1,0.3,0.5,0.7,0.9$, respectively from right (oblate) to left (prolate). The expected DM halo shapes from \protect\cite{Tenneti14} for $M_{200c} = 10^{12},10^{13},10^{14},10^{15} h^{-1} \textrm{M}_{\odot}$ at $z = 0.34$ are shown respectively from top right to lower left by white circles.}
\label{fig:e_qs}
\end{figure}

The trend above is in agreement with N-body simulations which predict that haloes are more prolate with increasing mass. Fig.~\ref{fig:e_qs} shows the predicted ellipticity, $e$, for triaxial DM haloes with axis ratios 1:$q$:$s$, as a function of $q$ and $s$, after projecting to 2D using \cite{Ryden1992} then averaging over random projection angles. Expected DM halo shapes from \cite{Tenneti14} for $M_{200c} = 10^{12},10^{13},10^{14},10^{15}\,h^{-1} \textrm{M}_{\odot}$ at $z = 0.34$ (same as DR7 LRGS) are shown in  Fig.~\ref{fig:e_qs} and overplotted in Fig.~\ref{fig:compare_e}. These predict a slope of $0.06$ per decade in mass, flatter than the value in equation~(\ref{eq:eslope}).

As discussed in Sec.~\ref{sec:misalign}, we do not expect the major axis of the stellar light to be perfectly aligned with the major axis of the projected DM halo. This leads to a lower ``effective''  ellipticity, $e\sbr{eff}$, measured by weak lensing. We consider two models for misalignment, both based on hydrodynamical simulations. \cite{Tenneti14} found low misalignment, leading to high $e\sbr{eff}/e$ ratios of 0.68 and 0.96 at $M_{\textrm{200c}} = 10^{12}$ and $3\times10^{13} h^{-1} M_{\odot}$, respectively. On the other hand, the typical misalignment measured by \cite{VelCacSch15} is larger, leading to lower values of $e\sbr{eff}/e$ than \cite{Tenneti14} (see Appendix \ref{sec:Vel15}). Both are shown in Fig.~\ref{fig:compare_e}. In both cases the alignment increases with increasing halo mass, leading to a steeper slope of $e_{\textrm{eff}}$ with halo mass, in better agreement with the observations. Overall, the \cite{Tenneti14} predictions are slightly, but not significantly, higher than the observational trend over the mass range covered by that study. While the \cite{VelCacSch15} predictions are, on average, lower than the observational fit. However, there are two important caveats regarding these comparisons. First, here we focus on red galaxies, whereas the results quoted above for simulations do not distinguish between red and blue galaxies. Second, \cite{VelCacSch15} uses the major axis of stars within the stellar half-mass radius.  Our major and minor axes are based on second moments above an isophotal threshold \citep[SExtractor]{BerArn96}. The position angles determined this way may be based on light that is more extended than those determined within the half-mass radius. \cite{VelCacSch15} show that the alignment improves when using stars at larger radii, so it is possible that if they had used a larger radius, the alignment would have been better and the ratio $e\sbr{eff}/e$ higher, bringing their predictions into better agreement. It is clear that future observational and theoretical studies will need to pay greater attention to measuring major axes and ellipticites in a more consistent way.

\section{Conclusions}
\label{sec:conclusion}
We studied the anisotropic lensing signal around LRGs from SDSS with source galaxies from an early internal shape measurement catalogue from UNIONS covering 1,500 square degrees. Fitting an NFW profile to the monopole shear profile of 18,000 DR7 LRGs yields an average mass of $M_{200c} \sim 2.7\times10^{13}$ M$_{\odot}/h$. When aligning our coordinate system with the major axis of the lens galaxy stellar light, we measure a mean halo ellipticity of $e=0.46\pm0.10$ and an aligned ellipticity ratio of $f_h=2.2\pm0.6$. This value is in agreement with other measurements of halo ellipticity from weak lensing \citep{ClaJai16,VanHoeJoa17}.  The 144,000 LRGs from BOSS are less massive, with an average mass of $M_{200c}=1.2\times 10^{13} M_{\odot}/h$. Repeating the analysis for the LRGs from BOSS, we found a mean halo ellipticity of $e=0.20\pm0.10$ and an aligned ellipticity ratio of $f_h=0.7\pm0.7$.

Combining our results together with previous measurements of halo ellipticity yields a trend with ellipticity increasing $0.10\pm0.06$ per decade in halo mass.

The prospects for improving halo ellipticity measurements from weak lensing are very promising. UNIONS is still underway, with the goal of covering $4,800$ square degrees with high quality multi-band imaging, which will yield photometric redshifts for source galaxies. The UNIONS survey area has large overlap with the footprint of the SDSS-based spectroscopic surveys, and, in the near future, the Dark Energy Spectroscopic Instrument \citep{DESICollaborationAghamousaAguilar2016}, which will allow larger and more comprehensive lens catalogues.

\section*{Acknowledgements}

We thank the referee for useful comments that improved this paper.

MJH acknowledges funding from an NSERC Discovery Grant. HH is supported by a DFG Heisenberg grant (Hi 1495/5-1), the DFG Collaborative Research Center SFB1491, as well as an ERC Consolidator Grant (No. 770935). 

This work is based on data obtained as part of the Canada-France Imaging Survey, a CFHT large program of the National Research Council of Canada and the French Centre National de la Recherche Scientifique. This work is additionally based on observations obtained with MegaPrime/MegaCam, a joint project of CFHT and CEA Saclay, at the Canada-France-Hawaii Telescope (CFHT) which is operated by the National Research Council (NRC) of Canada, the Institut National des Science de l’Univers (INSU) of the Centre National de la Recherche Scientifique (CNRS) of France, and the University of Hawaii. We are honored and grateful for the opportunity of observing the Universe from Maunakea, which has the cultural, historical and natural significance in Hawaii.  This research used the facilities of the Canadian Astronomy Data Centre operated by the National Research Council of Canada with the support of the Canadian Space Agency.

\section*{Data Availability}

A subset of the raw data underlying this article are publicly available via the Canadian Astronomical Data Center at \href{http://www.cadc-ccda.hia-iha.nrc-cnrc.gc.ca/en/megapipe/}{http://www.cadc-ccda.hia-iha.nrc-cnrc.gc.ca/en/megapipe/}. The remaining raw data and all processed data are available to members of the Canadian and French communities via reasonable requests to the principal investigators of the Canada-France Imaging Survey, Alan McConnachie and Jean-Charles Cuillandre. All data will be publicly available to the international community at the end of the proprietary period, scheduled for 2023.



\bibliographystyle{mnras}
\bibliography{refs}

\appendix
\section{Misalignments in Velliscig et al.\ 2015}
\label{sec:Vel15}

\cite{VelCacSch15} propose that the misalignment, $\theta\sbr{mis}$ between the projected major axis of the stellar light and that of the DM halo has a probability density that takes the form of the double Gaussian,
\begin{equation}
    P(\theta\sbr{mis}) = C \exp\left(-\frac{\theta\sbr{mis}^2}{2\sigma_1^2}\right) + D \exp\left(-\frac{\theta\sbr{mis}^2}{2\sigma_2^2}\right) + E \,.
\label{eq:double_gaussian}
\end{equation}
They tabulate parameters of the fit in their table B2. There is, however, a typographical error in that table. Consequently, we digitised their fig.~10 and refitted the parameters using the same functional form: the results are given in Table~\ref{tab:misalignment_parameters}. The LRG lenses studied in this paper are best described by the $13 < \log_{10}(M_{200c}/[$M$_{\odot}h^{-1}]) < 14$ halo mass bin.

\begin{table}
 \caption{Fit parameters for equation~(\ref{eq:double_gaussian}) that describes the probability density function of the misalignment between the projected stellar distribution and the projected total matter distribution, based on fig.~10 from \protect\cite{VelCacSch15}. The mass bin column gives the range in units $\log_{10}[M\sbr{200c}/ (h^{-1} \Msun)]$. $\theta\sbr{mean}$ and $\theta\sbr{med}$ are the mean and median misalignment angles, respectively. The probability density function is normalized to unity over the range $0^{\circ}$ to $90^{\circ}$.}
\label{tab:misalignment_parameters}
 \begin{tabular}{lcccccccc}
 \hline
mass & $\sigma_1$ & $\sigma_2$ & $C$ & $D$ & $E$ & $\theta\sbr{mean}$ & $\theta\sbr{med}$ & $e_{\textrm{eff}}/e$ \\
bin  & deg & deg &  deg$^{-1}$ &  deg$^{-1}$  &  deg$^{-1}$ & deg & deg &  \\
 \hline
$11-12$ & 5.1 & 28.3 & 0.0273 & 0.0085 & 0.0058 & 30 & 24 & 0.36 \\
$12-13$ & 5.1 & 31.6 & 0.0070 & 0.0118 & 0.0054 & 33 & 28 & 0.30 \\
$13-14$ & 14.8 & 32.9 & 0.0200 & 0.0086 & 0.0030 & 25 & 18 & 0.51 \\
$14-15$ & 9.6 & 25.9 & 0.0239 & 0.0142 & 0.0028 & 22 & 15 & 0.58 \\
\hline
\end{tabular}
\end{table}

\bsp	
\label{lastpage}
\end{document}